\documentclass[journal,10pt]{IEEEtran}

\usepackage[pdftex]{graphicx}
\usepackage{subfigure}
\usepackage[cmex10]{amsmath}
\usepackage{array}
\usepackage{cite}
\usepackage{amssymb}
\usepackage{caption}
\usepackage{url}
\usepackage{cite}
\usepackage{xcolor}

\usepackage{mathtools}
\DeclareMathOperator*{\argmax}{arg\,max}   

\graphicspath{{figures_v2/}}
\begin{document}

\title{Smart Jammer and LTE Network Strategies in An Infinite-Horizon Zero-Sum Repeated Game with Asymmetric and Incomplete Information}

\author{Farhan~M.~Aziz,
			Lichun~Li,
        	Jeff~S.~Shamma,
        	and~Gordon~L.~St\"{u}ber 

\thanks{F. M. Aziz is a systems engineer working on 4G/5G technologies at Intel Corporation, San Diego, CA and an alumnus of the Wireless Systems Laboratory (WSL), School of Electrical and Computer Engineering, Georgia Institute of Technology, Atlanta, GA, 30332, USA. G. L. St\"{u}ber is the professor and director of the Wireless Systems Laboratory (WSL). E-mail: \{faziz,stuber\}@ece.gatech.edu.}
\thanks{L. Li is an assistant professor with the Department of Industrial and Manufacturing Engineering, Florida A\&M University - Florida State University College of Engineering, Tallahassee, FL 32310, USA. E-mail: lichunli@eng.famu.fsu.edu.}
\thanks{J. S. Shamma is the professor and director  of the RISC Lab, Computer, Electrical and Mathematical Science and Engineering (CEMSE) Division, King Abdullah University of Science and Technology (KAUST), Thuwal, Saudi Arabia. E-mail: jeff.shamma@kaust.edu.sa.}

}

{}

\maketitle

\begin{abstract}
LTE/LTE-Advanced networks are known to be vulnerable to denial-of-service (DOS) and loss-of-service attacks from smart jammers. In this article, the interaction between a smart jammer and LTE network (eNode~B) is modeled as an infinite-horizon, zero-sum, repeated game, with asymmetric and incomplete information. The smart jammer and eNode~B are modeled as the informed and the uninformed player, respectively. The main purpose of this article is to construct efficient suboptimal strategies for both players that can be used to solve the above-mentioned infinite-horizon repeated game with asymmetric and incomplete information. It has been shown in game-theoretic literature that security strategies provide optimal solution in zero-sum games. It is also shown that both players' security strategies in an infinite-horizon asymmetric game depend only on the history of the informed player's actions. However, fixed-sized sufficient statistics are needed for both players to solve the above-mentioned game efficiently. The smart jammer (informed player) uses its evolving belief state as the fixed-sized sufficient statistics for the repeated game. Whereas, the LTE network (uninformed player) uses worst-case regret of its security strategy and its anti-discounted update as the fixed-sized sufficient statistics. Although fixed-sized sufficient statistics are employed by both players, optimal security strategy computation in $\lambda$-discounted asymmetric games is still hard to perform because of non-convexity. Hence, the problem is convexified in this  article by devising ``approximated'' security strategies for both players that are based on approximated optimal game value. However, ``approximated'' strategies require full monitoring. Therefore, a  simplistic yet effective ``expected'' strategy is also constructed for the LTE network (uninformed player) that does not require full monitoring. The simulation results show that the smart jammer plays non-revealing and misleading strategies against the network for its own long-term advantage.
\end{abstract}


\IEEEpeerreviewmaketitle

\section{Introduction}

\IEEEPARstart{L}{TE/LTE-A} (\cite{3GPP, LTE}) networks have been deployed  around the world providing advanced data, Voice-over-LTE (VoLTE),  multimedia and location-based services to more than 3.2 billion subscribers via 681 commercial networks \cite{GSA}. However, it has been previously shown that Long Term Evolution (LTE) and LTE-Advanced (LTE-A) networks are vulnerable to control-channel jamming attacks from  \textit{smart jammers} who can ``learn'' network parameters and ``synchronize'' themselves with the network even when they are not attached to it (cf. \cite{Jover_2014, Aziz_2014, Reed_2015, Aziz_2015, Reed_2016, Aziz_TVT}). It is shown in the above-referenced articles that such a \textit{smart jammer} can launch very effective \textit{denial-of-service (DOS)} and \textit{loss of service} attacks without even hacking the network or its components. Recently, Garnaev and  Trappe \cite{Garnaev_2018} also looked into the possibility when the rival is not smart in a jamming game with incomplete information. Hence, pursuing autonomous techniques to address potentially devastating wireless jamming problem has become an active research topic.

In this article, the interaction between the LTE network and the smart  jammer is modeled as an \textbf{infinite-horizon zero-sum repeated~\footnote{A \textit{repeated game} results when an underlying stage game is played over many stages, and a player may take into account its observations about all previous stages when making a decision at each stage \cite{Handbook_v1}.} Bayesian game} with \textbf{asymmetric} and \textbf{incomplete information}. This article is similar to previously published articles by the authors (\cite{Aziz_2014, Aziz_2015, Aziz_TVT}) with the exception of zero-sum and $\lambda$-discounted utility. The main purpose of this article is to construct efficient suboptimal strategies for both players to solve the above-mentioned infinite-horizon game with asymmetric and incomplete information. Asymmetric information games (cf. \cite{Handbook_v1} - \cite{Repeated_Games}) provide a rich framework to model situations in which one player lacks complete knowledge about the ``state of the nature''. The player who possesses complete knowledge about the state of the nature is known as the informed player and the one who lacks this knowledge is called the uninformed player. The  \textbf{smart jammer} is modeled as the \textbf{informed row player}, whereas LTE \textbf{eNode~B} is modeled as the \textbf{uninformed column player}. The informed player deals with the ultimate and subtle tradeoff of exploiting its superior information at the cost of revealing that information via its actions or some other (unavoidable) signals during repeated interactions with other players (cf. \cite{Handbook_v1, Incomplete_Information}).  In most game-theoretic literature on repeated games with asymmetric information, the informed player's strategy is computed based on how much information it should reveal for an optimal or suboptimal policy. Furthermore, many informed player zero-sum formulations model the uninformed player as a Bayesian player in order to solve asymmetric games (cf. \cite{Malachi_2012} - \cite{Lichun_2017}). However, relatively little work has been done  to address the  optimal strategy computation of the uninformed player in an infinite-horizon repeated zero-sum game with asymmetric information \cite{Kamble}. The main difficulty arises from the fact that the uninformed player lacks complete knowledge about the state of nature and informed player's belief state, which plays a crucial role in determining players' payoffs and strategies. However, it has been shown in\cite{Zero_Sum} that the uninformed player's security strategy does exist in finite-horizon and infinite-horizon games with discounted and bounded average cost formulations. Furthermore, it has been shown in \cite{OR} that the uninformed player's security strategy does not depend on history of its own actions.

This article attempts to solve the above-mentioned LTE network vs. smart jammer game by constructing efficient linear programming (LP) formulations for both players' ``approximated'' strategy computation and unique ``expected'' strategy computation for the eNode~B. The informed player's (the smart jammer) security strategy (optimal strategy in the worst-case scenario) only depends on the history of its own actions and is independent of the other player's actions. The smart jammer models the uninformed player as a Bayesian player, making Bayesian updates with an evolving belief state. However, in order to solve the infinite-horizon game efficiently, fixed-sized sufficient statistics are needed for both players that do not grow with the horizon. The evolving belief state serves as sufficient statistics for the informed player in a $\lambda$-discounted asymmetric repeated game. On the other hand, the uninformed player's (the eNode~B) security strategy does not depend on the history of its own actions, but rather depends on the history of the informed player's actions. However, the uninformed player does not have access to the informed player's belief state and needs to find different fixed-sized sufficient statistics. Fortunately, the uninformed player's security strategy in the dual game depends only on a fixed-sized sufficient statistics that is fully available to it. Furthermore, the uninformed player's security strategy in the dual game, with initial worst-case regret vector, also serves as its security strategy in the primal game. Therefore, initial worst-case regret of its security strategy and its anti-discounted update (which is the same size as the cardinality of system state) is used as the fixed-sized sufficient statistics for the uninformed player. Although the above-mentioned sufficient statistics are fixed-sized for both players in an infinite-horizon game, the optimal security strategy computation in $\lambda$-discounted asymmetric game is still hard to compute because of non-convexity \cite{Sandholm}. Consequently, ``approximated'' security strategies based on an approximated optimal game value with guaranteed performance are computed for both players that are based on a recent work by Li and Shamma \cite{Lichun_2017}.  

The above-mentioned ``approximated'' security  strategies require \textit{full monitoring}~\footnote{\textit{Full monitoring} requires that all players are capable of observing previous actions of their opponents with certainty after each stage \cite{Handbook_v1}.}. Since eNode~B cannot observe smart jammer's actions with complete certainty, a unique ``expected'' strategy formulation is also presented for the uninformed player (the eNode~B) that does not require full monitoring. Both the smart jammer and the LTE eNode~B exploit the ``approximated'' and ``expected'' formulations to compute their sub-optimal yet efficient strategies in order to maximize their corresponding utilities. The main idea of the zero-sum formulation is to find an (almost) optimal yet tractable strategy for both players in the worst-case scenario. It is to be noted here that the smart jammer is the maximizer  in the zero-sum game and  the eNode~B is the minimizer.  Also, full monitoring is only limited to observing actions of the opponent  and, hence, does not reveal any private information of ordinary UEs in the network to the smart jammer. 

\subsection{Related Work}
Game theory (cf. \cite{Game_Theory_2, Handbook_v1, Game_Theory, Incomplete_Information, Zero_Sum, Repeated_Games, Handbook_v4}) provides a rich set of mathematical tools to analyze and address conflict and cooperation scenarios in multi-player situations, and as such has been applied to a multitude of real-world situations in economics, biology, cyber security, multi-agent networks, wireless networks (cf. \cite{Mackenzie, Altman, Walid_Saad}) and more. In this article, the interaction between the LTE network and the \textit{smart jammer} is modeled as an \textit{infinite-horizon zero-sum repeated game with asymmetric information}. 

Zero-sum repeated game formulations have been studied extensively in the game-theoretic literature, including asymmetric information cases, such as Chapter 5 of \cite{Handbook_v1}, Chapter 4 of \cite{Incomplete_Information}, Chapters 2 - 4 of \cite{Zero_Sum}, and Chapter 2 of \cite{Handbook_v4}. However, most of the prior work on asymmetric zero-sum repeated games deals with the informed player's viewpoint. For example, \cite{Handbook_v1} and \cite{Incomplete_Information} pointed out that the informed player might reveal its superior information implicitly by its  actions and, hence, may want to refrain from certain actions in order not to reveal that information. In case of full monitoring, playing non-revealing strategy for the informed player is equivalent to not using its superior information \cite{Handbook_v1}. Furthermore, \cite{Zero_Sum} showed that the informed player's belief state (conditional probability of the game being played given history of informed player's actions) is its sufficient statistics to make long-run decisions. Hence, many informed player's strategies (cf. \cite{Malachi_2012} - \cite{Lichun_2017}) use  the belief state as their sufficient statistics. However, \cite{Sandholm} showed that computing the optimal value of the infinite-horizon repeated game is non-convex and  identified computational complexities involved in solving infinite-horizon games. Therefore, the above-mentioned articles approximate the optimal game value via linear programming. On the other hand, limited work has been done for the uninformed player's optimal strategy computation as compared to the vast research done for the informed player \cite{Kamble}. It is, however, known that the uninformed player's security strategy exists in infinite-horizon repeated zero-sum games, and that it does not depend on the history of its own actions (cf. \cite{Lichun_2016, OR}). But, efficient computation of the uninformed player's optimal security strategy is still an open problem. Recently, \cite{Kamble} suggested that  the uninformed player could use its expected payoff for each candidate game as sufficient statistics since it is unaware of the game being played. Similarly,  \cite{Lichun_2016} used realized vector payoff as the uninformed player's sufficient  statistics to compute its efficient but suboptimal strategy in finite-horizon zero-sum repeated games. However, it is to be noted here that all of these formulations are based on commonly-assumed notion of full monitoring in which players can  perfectly observe their opponent's  actions. This article also utilizes the notion of full monitoring for its ``approximated'' strategy computation. 

Although there has been quite a lot of work done on infinite-horizon repeated zero-sum games with asymmetric information, there does not exist any tractable non-zero-sum formulations for the uninformed player that can be used for its optimal strategy computation in infinite-horizon asymmetric repeated games. Most of the classic general-sum (non-zero-sum) game-theoretic literature like Chapter 6 of \cite{Handbook_v1} and Chapters V and IX of \cite{Repeated_Games} focus on the characterization and existence of equilibria in repeated games with asymmetric information, and deal with the optimal strategy construction for the  full monitoring case. Chapter V of \cite{Repeated_Games} also suggests using \textit{approachability theory} for the construction of the uninformed player's strategy for the full monitoring case. However, none of these formulations result in  efficient computation of the uninformed player's optimal strategy. This problem gets further complicated for general-sum (non-zero-sum) games with imperfect monitoring. For example, \cite{R1-2} pointed out that the solution of a general-sum (non-zero-sum) stochastic game with both incomplete knowledge and imperfect monitoring is an open problem and there is no well-established solution available so far. To the best of our knowledge, that is still the case for repeated as well as stochastic general-sum (non-zero-sum) games (e.g., see \cite{Kamble, Aziz_Dissertation}). This is one of the main reasons that the LTE network and smart jammer interaction is not modeled as a general-sum (non-zero-sum) game in this paper.  

Bayesian approaches have been widely used to solve asymmetric information problems in which updated belief state can be used as sufficient statistics for the informed player. Belief state often serves as  a tool for updating the internal notion of a player's knowledge related to another. For example, \cite{Malachi_2012} - \cite{Lichun_2017} modeled the uninformed player as a Bayesian player in order to compute the informed player's suboptimal strategies in repeated zero-sum games. Similarly, \cite{Garnaev_2014} - \cite{Garnaev_2016} used Bayesian approaches to devise an uninformed player's strategy based on expected payoff, and \cite{R1-2} employed Bayesian Nash-Q learning in  an incomplete information stochastic game and used Bayes' formula to update belief  of an  Intrusion Detection System (IDS). Another commonly used technique to address lack of information problems is state estimation. For example, \cite{Gursoy_2014} used a Kalman filter to estimate the state of an observable, linear, stochastic dynamic system in an infrastructure security game. However, the system of interest and game dynamics in this article are nonlinear and may not be completely-observable. Therefore, the applicability of state estimation techniques is very limited.

To the best of our knowledge, there does not exist any explicit formulations for the optimal strategy computation of the uninformed player in an infinite-horizon repeated zero-sum game with asymmetric information \cite{Kamble}. However, it has been shown in\cite{Zero_Sum} that the uninformed player's security strategy does exist in finite-horizon and infinite-horizon games with discounted and bounded average cost formulations. Moreover, it has been shown in \cite{OR} that the uninformed player's security strategy does not depend on history of its own actions. Nevertheless, a recent  LP formulation in \cite{Lichun_2017} provides an efficient technique for the explicit ``approximated'' strategy computation of the uninformed player in infinite-horizon asymmetric repeated zero-sum games, but with the assumption of full monitoring.  

Furthermore, there are multiple differences between this article and \cite{Aziz_TVT} which is focused only on estimating the jammer type at the beginning of the game and as such strategies computed in that article cannot be used for long-term interaction. On the other hand, this article is focused on computing both players' strategies for very long-term interaction. Moreover, smart jammer in \cite{Aziz_TVT} is modeled as a myopic player (i.e., it only cares about short-term utility) as opposed to being modeled as a strategic player in this article. In addition, the interaction between the smart jammer and the eNode~B in \cite{Aziz_TVT} is modeled as a general-sum (non-zero-sum) game without perfect monitoring  as opposed to zero-sum formulation in this article. The eNode~B exploits ``jamming sense'' part of the algorithm presented in \cite{Aziz_TVT} to invoke strategy computation algorithms discussed in this article. On the other hand, the article \cite{Lichun_2017} written by co-authors Li and Shamma is focused on an approximated yet efficient LP formulation for both players' strategies in an abstract repeated zero-sum game with perfect monitoring. This ``approximated'' strategy construction technique is further extended to realistic smart jammer and LTE network interaction in this article. On top of that, a unique ``expected'' strategy formulation is also explored in the article. 
 
Smart jamming problem in LTE networks has been studied extensively lately. However, to the best of our knowledge,   none of the articles published so far studied the smart jamming problem in  LTE networks in a game-theoretic manner.

\section{Smart Jamming in LTE Networks}
Potential \textit{smart jamming} attacks and suggested network countermeasures are the same as described in \cite{Aziz_2014}. They are briefly discussed here for the sake of completeness. 

\subsection{Smart Jamming Attacks on an LTE Network}
The set of \textit{smart jammer's} pure actions consists of the following jamming attacks~\footnote{See \cite{3GPP} or \cite{LTE} for the description of various LTE channels}: 

\begin{enumerate}
\item 	\textbf{$a_j^1$ =  Inactive \textit{(no jamming)}}: corresponds to default UE operation when no jammer is active in the network.

\item 	\textbf{$a_j^2$ = Jam \textit{CS-RS}}: corresponds to OFDM pilot jamming/nulling extensively studied in the literature. This action prevents all UEs from demodulating data channels by prohibiting them from performing coherent demodulation, degrades cell quality measurements and blocks initial cell acquisition in the jamming area.

\item	\textbf{$a_j^3$  = Jam \textit{CS-RS + PUCCH}}: corresponds to jamming PUCCH in the UL in addition to DL CS-RS jamming. This action could be more catastrophic for Connected mode UEs as compared to jamming the CS-RS alone due to eNode~B's loss of critical UL control information but requires more sophisticated dual-band jamming. 

\item 	\textbf{$a_j^4$ = Jam \textit{CS-RS + PBCH + PRACH}}: corresponds to jamming DL broadcast channel PBCH and UL random access channel PRACH in addition to pilot nulling. This action is intended to block reselection/handover of UEs from neighboring cells and block synchronization of idle mode and out-of-sync UEs.

\item	\textbf{$a_j^5$ = Jam \textit{CS-RS + PCFICH + PUCCH + PRACH}}: corresponds to jamming CS-RS and PCFICH in the DL and jamming PUCCH and PRACH in the UL. This action is intended to cause loss of DL and  UL grants, radio link failures and  loss of UE synchronization mostly in Connected mode UEs.\\
\end{enumerate}

Although jamming individual control channels may also cause denial-of-service (DOS) attacks, jamming effects may not have specific consequences desired by the smart jammer. For example, jamming CS-RS alone could be limited to specific part of a cell depending on the jammer location and its transmit power  and jamming PBCH alone may only prevent a small fraction of UEs to not reselect/handover to the cell if they have not visited that particular cell recently. Even though concatenation of jamming multiple control channels requires distribution of the jammer's transmit power among all jamming activities, it allows the smart jammer to target specific aspects of network operation, such as cell reselection/handover, data download and upload etc. 

In addition to the above-mentioned pure actions, the \textit{smart jammer} uses its probability of jamming ($p_j$) and transmit power ($P_j$) to decide  when to jam the network and how much power to use during a particular jamming attack. The duty cycle of each action is also implicitly modeled in the utility function of the jammer. Thus, the smart jammer launches \textit{denial-of-service (DoS)} and \textit{loss of service} attacks on the LTE network by employing these actions and can be easily implemented using a software-defined radio (SDR) and a colluding UE.

\subsection{Suggested Network Countermeasures}
It is proposed that the network can use the following (pure) countermeasures in case of a jamming attack: 

\begin{enumerate}
\item  \textbf{$a_0^1$ = Normal \textit{(default action)}}: corresponds to default network operation.

\item 	\textbf{$a_0^2$ = Increase \textit{CS-RS} Transmit Power}: corresponds to \textit{pilot boosting} in order to alleviate CS-RS jamming at the expense of transmitting other channels at lower transmit power than normal operation.

\item 	\textbf{$a_0^3$ = \textit{Throttle}}: corresponds to a specific \textit{threat mechanism} when all active UEs' DL/UL grants (and hence throughputs) are throttled.

\item 	\textbf{$a_0^4$ = Change \textit{eNode~B $f_c$ + SIB~2}}: corresponds to a specific \textit{interference avoidance mechanism} when the network ``relocates'' its carrier frequency $f_c$ to a different carrier within its allocated band/channel and rearranges itself into a lower occupied bandwidth configuration. It also changes its PRACH configuration parameters in SIB~2 to alleviate PRACH jamming.

\item 	\textbf{$a_0^5$ = Change eNode~B \textit{Timing}}: corresponds to a specific \textit{interference avoidance mechanism} in which network ``resets'' its frame/subframe/slot/symbol timing and SIB 2 parameters. Ongoing data sessions would be handed over to neighboring cells before the ``reset'' and the cell would not be available during transition. \\
\end{enumerate}

It is to be noted here that the above-mentioned countermeasures do not require any exogenous information or significant changes in 3GPP specifications and can be implemented easily with current technology. Furthermore, the network is not aware of jammer's location, jamming waveform, and its probability of jamming $p_j$. Also, the average duty cycle and eNode~B's transmit power ($P_0$) determine the power consumption of  the network, modeled in its utility function. The curious reader is encouraged to see \cite{Aziz_TVT, Aziz_Dissertation} for further details on smart jammer actions and network countermeasures.

\section{LTE Network \& Smart Jammer Dynamics}
\subsection{Network Model}
The network model used in this article is the same as developed in \cite{Aziz_TVT} and is briefly discussed here for the sake of completeness.

UEs arrive in the cell according to a \textit{homogeneous 2D Stationary Spatial Poisson Point Process (SPPP)} with the rate $\Lambda$ per unit area and are \textit{uniformly distributed} over the entire cell conditioned on the total number of users \textit{N}. The large-scale path loss is modeled using the \textit{Simplified Path Loss Model} \cite{Goldsmith}.

\begin{equation} \label{eq: path_loss}
P_r (dBm) 	=	P_t (dBm) + K (dB) - 10\gamma \mathrm{log_{10}}\left(\frac{d}{d_0}\right)
\end{equation}
where
$P_r$ is the received power, $P_t$ is the transmitted power, $K (dB) = 	20 \mathrm{log_{10}} \left(\frac{\lambda}{4 \pi d_0}\right)$ is a constant, $\gamma$ is the \textit{path loss exponent}, $d$ is the distance between the transmitter and receiver, and $d_0$ is the outdoor reference distance for antenna far field. The small-scale multipath fading is modeled using \textit{exponentially distributed Rayleigh-faded} channel gains at each subcarrier. Thus, the instantaneous SINR $\Gamma[k]$ of a particular OFDM subcarrier $k$ is modeled as follows:

\begin{equation} \label{eq: SINR_1}
\Gamma[k] 		= 	\frac{P_0[k] |h|^2 K (\frac{R_0}{d_0})^{-\gamma}}{\sigma^2 + P_j[k] |g|^2 K (\frac{R_j}{d_0})^{-\gamma} }
\end{equation}
where
$P_0$ and $P_j$ are desired and jammer transmit powers, $|h|^2$ and $|g|^2$ are \textit{exponentially distributed Rayleigh-faded} channel gains, $R_0$ and $R_j$ are large-scale distances from desired transmitter and jammer respectively, $\gamma$ is the path loss exponent, and $\sigma^2$ is the noise variance at the receiver. It is assumed that \textit{Inter-Cell Interference (ICI)} is independent of jamming and, hence, any residual ICI can be lumped together in the noise variance $\sigma^2$ for the scope of this article. It is further assumed that $\sigma^2$ is the same at all receivers.

The SINR in (\ref{eq: SINR_1}) can be re-written in terms of the \textit{Carrier-to-Jammer ratio} $C/J$ as follows:

\begin{equation} \label{eq: SINR_2}
\Gamma[k] 		= 	\frac{(\frac{C}{J}) |h|^2 K (\frac{R_0}{d_0})^{-\gamma}}{(\frac{\sigma^2}{P_j[k]}) + |g|^2 K (\frac{R_j}{d_0})^{-\gamma} }
\end{equation}

Equations (\ref{eq: SINR_1}) and (\ref{eq: SINR_2}) are used to model the SINR of narrowband flat-faded signals and channels like CS-RS, PCFICH, PUCCH etc.. However, wideband channels like PDSCH and PUSCH cannot be  modeled using (\ref{eq: SINR_1}) or (\ref{eq: SINR_2}). Furthermore, SINR estimation is done in the frequency domain.

In addition, the LTE network's $m$th user's DL PDSCH throughput $\mathcal{R}_m (k,l)$ in the $k$th resource block during the $l$th subframe is modeled as a fraction $\epsilon \in (0, 1]$ of \textit{Shannon's AWGN Channel Capacity} as described in (\ref{eq: throughput}).

\begin{equation} \label{eq: throughput}
\mathcal{R}_m (k,l) 	= 	\epsilon \; \mathbb{W}_{\text{RB}} \; \mathrm{log_2} [1 + \Gamma_m^{\text{PDSCH}} (k,l)]
\end{equation}
where $\mathbb{W}_{\text{RB}}$ is the bandwidth of a single RB i.e. 180 kHz. For the purposes of this article, it is assumed that  $\epsilon = 1$.

The $m$th user's total throughput in a given subframe is the sum of its assigned RBs' throughput  for that particular subframe. It is modeled that the eNode~B uses a Proportional Fair Scheduling (PFS) \cite{PFS} algorithm to allocate resources to its users. User $m$ is allocated in resource block $k$ during the $l$th subframe if the ratio of its achievable instantaneous data rate and long-term average throughput in (\ref{eq: PFS}) is the highest among all the active users in the network. The long-term average throughput of user $m$, $\overline{\mathcal{R}}_m (l)$ during subframe $l$ is computed using the recursive equations (\ref{eq: PFS}) and (\ref{eq: PFS_avg_tput}) below:

\begin{equation} \label{eq: PFS}
\hat{m}_k = \argmax_{m' = 1,...,N} \left\lbrace \frac{\mathcal{R}_{m'}(k,l)} {\overline{\mathcal{R}}_{m'}(l)} \right\rbrace
\end{equation}

\begin{equation} \label{eq: PFS_avg_tput}
\overline{\mathcal{R}}_m (l) = \left(1 - \frac{1}{t_c}\right) \overline{\mathcal{R}}_m (l-1) + \frac{1}{t_c} \sum_{k=1}^K \mathcal{R}_m (k,l) \mathcal{I}(\hat{m}_k = m)
\end{equation}
where $t_c$ represents fairness time window, and $\mathcal{I}$ is an indicator function.

In general, the overall LTE network dynamics can be modeled as a \textit{highly nonlinear dynamical system} described by:

\begin{equation} \label{eq: utility_function}
\chi^+ = f(\chi,\theta,a^0,a^j,\omega)
\end{equation}
where
$\chi \in \mathbb{R}^{M \times K}$ represents state of the network (not to be confused with the game-theoretic state of nature $\theta$) with each row corresponding to the user $m \in M$, including $K$ elements for each user (such as, SINRs $\Gamma_m$ of its control and data channels, and average throughput for user $m \in M$); $\theta$ represents the game-theoretic state of nature (jammer type) described in the next section; $a^0 \in \mathcal{A}_0$ represents eNode~B action; $a^j \in \mathcal{A}_j$ represents jammer's action and $\omega$ characterizes the randomness in the network induced by the channel, arbitrary user locations, varying transmit power levels, PFS scheduling and other sources of randomness in the network. Thus, the network dynamics are modeled as a \textit{Partially-Observable Markov Decision Process (POMDP)}. 

Evidently, it is nontrivial and intractable to model the LTE network and smart jammer dynamics analytically. Hence, these abstracted dynamics are simulated in \textit{MATLAB} without losing any modeling fidelity. Although this article can be used as a building block for more complicated scenarios, multi-cell and multi-jammer scenarios are beyond the scope of this article. 

\subsection{Game-Theoretic Model}
\textbf{Notations:} For the rest of this paper, the following mathematical notations are used. Let $\mathcal{A}$ be a finite set. Its cardinality is denoted by $|\mathcal{A}|$, and the set of all probability distributions over $\mathcal{A}$ is indicated by $\Delta(\mathcal{A})$.\\

\subsubsection{Game Model}
The interaction between the LTE network and the \textit{smart jammer} is modeled as a strictly competitive \textbf{infinite-horizon zero-sum repeated Bayesian game $\mathcal{G}$ with asymmetric and incomplete information}, with the \textit{smart jammer} as the informed (row) player and the eNode~B as the uninformed (column) player. Infinite-horizon games are used to model situations in which horizon length is not fixed in advance and there is a non-zero probability at the end of each stage that the game will continue for next stage (e.g., see \cite{Game_Theory}). The game  $\mathcal{G}$ is described by
\begin{itemize}
\item 	$\mathcal{N}$ = \{smart jammer, eNode~B\}, the set of players,
\item 	$\Theta$, the set of states of nature (jammer types),
\item 	$p_0 \in \Delta (\Theta)$, the prior probability distribution on $\Theta$, which is common knowledge,
\item 	$\mathcal{A}_j$ and $\mathcal{A}_0$, the set of pure actions of the \textit{smart jammer} and the eNode~B, respectively as described in Section II, where $a^j \in \mathcal{A}_j$ and $a^0 \in \mathcal{A}_0$ represent corresponding elements in these sets,
\item 	$\mathcal{H}$, a set of sequences such that each $H \in \mathcal{H}$ is a history of observations,
\item 	$\mathcal{I}_i$, the information partition of player $i$ and
\item 	$\mathcal{U}_i \colon \Theta \times \mathcal{A}_j \times \mathcal{A}_0 \rightarrow \mathbb{R}$, the single-stage utility function of player $i$, $\mathcal{U}_i^\theta\in\mathbb{R}^{|\mathcal{A}_j|\times |\mathcal{A}_0|}$ the utility matrix of player $i$ given jammer type $\theta$ whose element $\mathcal{U}_i^\theta(a^j,a^0)$ is $\mathcal{U}_i(\theta,a^j,a^0)$, $\forall a^j \in\mathcal{A}_j, a^0 \in\mathcal{A}_0$. \\
\end{itemize}

Following the convention used in game-theoretic literature including \cite{Lichun_2017}, the informed player, i.e., the \textbf{smart jammer} is played as the \textbf{maximizer (row player)}, whereas the uninformed player, i.e., the \textbf{eNode~B} is played as the \textbf{minimizer (column player)}. It is to be noted here that ordinary user equipments (UEs) are not modeled as players in this article.\\

\subsubsection{Jammer Types}
The type  $\theta \in \Theta$ of \textit{smart jammer} is classified as:
\begin{itemize}
\item 	\textit{Type I:}  	\textbf{Cheater}
\item 	\textit{Type II:}  	\textbf{Saboteur}\\
\end{itemize}

The type \textit{Cheater} is used  to model the jammer  with the intent of getting more resources for itself as a result of reduced competition among UEs. Thus, a cheating UE is always present in the network with an active data session. On the other hand, the type \textit{Saboteur} is used to model the  jammer with the intent of causing highest possible damage to the network resources. Thus, a sabotaging UE may be unattached to the network. It is to be noted here that the \textit{``Normal (inactive)''} jammer type is not modeled here because jammer is not present in that state and it is in the best interest of the network to play default normal action in that case. The strategy algorithms presented in this article are invoked when jammer is present in the network and/or when jamming is sensed by the network using ``jamming sense'' part of the algorithm presented in \cite{Aziz_TVT}. It is to be noted here that absence of jamming does not imply jammer's absence as an active smart jammer may also decide to play \textit{``inactive''} action as part of its strategy during an attack. \\

\subsubsection{Strategies}
Both the network and the jammer are modeled as rational and strategic. By definition, a \textit{pure strategy} of a player is a mapping from each non-terminal history to a pure action and a \textit{mixed strategy} is a probability measure $\Delta$ over the set of its pure strategies. A \textit{behavioral strategy} specifies a probability measure $\Delta$ over its available actions at each stage when an action needs to be taken \cite{Game_Theory}. Also, the \textit{best response (BR)} is the strategy (or strategies) that produces the most  favorable outcome for a player given  other players' strategies \cite{Game_Theory_2}. Two types of suboptimal security strategies for infinite-horizon game are presented in this article, as discussed in Section V.\\

\subsubsection{Information Partitions}
The jammer is informed of its own type $\theta$. However, eNode~B is only informed about the prior probability distribution $p_0 \in \Delta(\Theta)$. This results in a \textit{game with asymmetric information}, with lack of information on the network side, making eNode~B the uninformed player.\\

\subsubsection{Observable Signals}
It is assumed that for the ``approximated'' strategy computation, players can observe each other's actions with certainty after each stage, i.e. full monitoring requirements are  satisfied. This is a very widely used assumption used in classic and modern game-theoretic literature (e.g., Chapter 6 of \cite{Handbook_v1}). The network can distinguish between smart jammer's different actions at high SNR and can make reasonable estimates  at low SNR. However, the imperfect monitoring case is beyond the scope of the ``approximated'' formulation presented in this article. On the other hand, the ``expected'' strategy formulation for repeated games does not require any full monitoring.\\

\subsubsection{Utilities}
Both players' utility functions are based on their key performance indicators (KPIs)  and are defined to reflect a strictly competitive (zero-sum) setting, i.e., one player's gain is the other player's loss as described by:

\begin{equation} \label{eq: zero_sum_utility}
\mathcal{U}_0 		=		-\mathcal{U}_j
\end{equation}

When the system state is \textit{Cheater}, the zero-sum utility function is simplified as

\begin{equation} \label{eq: Uc}
\begin{split}
\mathcal{U}_j^c (a^j,a^0) 	& = 	\alpha^{\mathcal{R}_c} \mathbb{E}_w \Big[ \delta(\mathcal{R}_c^{\text{norm}}) (a^j,a^0) \Big]	\\
&  -\alpha^{\mathcal{N}_c} \mathbb{E}_w \Big[ \mathcal{N}_c^{\text{norm}} (a^j,a^0)  \Big]
\end{split}
\end{equation}
where $\delta(\mathcal{R}_c^{\text{norm}})$ represents change in the \textit{Cheater}'s normalized average throughput from the baseline scenario, $\alpha^{\mathcal{R}_c}$ represents its corresponding weight, $\mathcal{N}_c^{\text{norm}}$ represents the normalized average number of  Connected mode UEs in the network when the \textit{Cheater} is present, $\alpha^{\mathcal{N}_c}$ represents its corresponding weight and $\mathbb{E}_w$ represents expectation with respect to randomness caused by $w$ as mentioned in (\ref{eq: utility_function}).

The \textit{Cheater} tries to maximize (\ref{eq: Uc}) in order to increase its throughput from the baseline scenario and reduce the number of Connected mode UEs in the network which at the same time reduces the competition for limited network resources. The eNode~B, on the other hand, tries to minimize (\ref{eq: Uc}) to do the opposite, hence, creating a proper zero-sum game.

Similarly, the zero-sum utility function is defined in (\ref{eq: Us}) when the system state is \textit{Saboteur}.

\begin{equation} \label{eq: Us}
\begin{split}
\mathcal{U}_j^s (a^j,a^0) 	&	=		-\alpha^{\mathcal{N}_s} \mathbb{E}_w \Big[ \mathcal{N}_s^{\text{norm}} (a^j,a^0) \Big] \\
&	- \alpha^{\mathcal{R}_{\text{eNB}}} \mathbb{E}_w \Big[ \mathcal{R}_{\text{eNB}}^{\text{norm}} (a^j,a^0) \Big]
\end{split}
\end{equation}
where $\mathcal{N}_s^{\text{norm}}$ represents the normalized average number of  Connected mode UEs in the network when \textit{Saboteur} is present, $\alpha^{\mathcal{N}_s}$ represents its corresponding weight, $\mathcal{R}_{\text{eNB}}^{\text{norm}}$ represents eNode~B's normalized average throughput/UE, $\alpha^{\mathcal{R}_{\text{eNB}}}$ represents its corresponding weight and $\mathbb{E}_w$, again, represents the expectation with respect to randomness caused by $w$ as mentioned above.

The \textit{Saboteur} tries to maximize the opposite (negative of) eNode~B utility defined in terms of average number of Connected mode users and average throughput/UE, hence, defining the zero-sum game.

Note that there are no ``unilateral'' fixed costs associated with either player in the above-mentioned zero-sum construction. This means that the game would be played without modeling higher ``fidelity'' parameters like players' duty cycles and implicit cost associated with eNode~B actions like \textit{'f Change'} and \textit{'Timing Change'}. However, this ``fidelity'' loss does not affect the inherent nature of smart jammer and  network interaction and, hence, can be discounted. It is also to be noted here that the utility functions are different for different jammer type, which is a common phenomenon in Bayesian games. Furthermore, the key performance indicators (KPIs) are functions of observable parameters only, for example, eNode~B's utility is a function of parameters observed from \textit{Connected Mode} UEs. 

Interestingly, none of the players need to compute their utilities explicitly as it is not used to make strategy decisions in repeated games as discussed later in Section V. The payoffs are received by both players as a result of the interaction between the smart jammer and the network. Even though the network does not know the jammer type, it can compute expected utility for minimization based on the prior (and updated belief) probability of specific jammer type presence.\\

\subsubsection{Game Play}
At the beginning of the game, nature flips a coin and selects $\theta \in \Theta$ (jammer type) according to $p_0 \in \Delta(\Theta)$, which remains fixed for the rest of the game. The jammer is informed about its selected type but eNode~B is not. However, in a \textit{repeated game}, eNode~B's history of interaction with the jammer evolves with time which may affect its belief about $\theta$.

\section{Single-Shot Game}
The single-shot game is played between the \textit{smart jammer} as the \textbf{maximizer} (row player) and the network as the \textbf{minimizer} (column player). The \textbf{maxmin} value for the row player for given state $\theta$ is denoted by $\underline{v}$; whereas the \textbf{minmax} value for the column player is denoted by $\overline{v}$. It is widely known that $\underline{v} \leq \overline{v}$ is always true. However, when $\underline{v} = \overline{v}$ is satisfied, then  the game is said to have a \textbf{value} $v = \underline{v} = \overline{v}$. The legendary \textit{von Neumann's} celebrated \textbf{Minmax Theorem} states that any matrix game has a value $v$ in mixed strategies and the players have optimal strategies \cite{Zero_Sum}, i.e., the  \textit{minmax solution} of a zero-sum game is the same as the \textit{Nash equilibrium}. Both players play their \textbf{security strategies} in a zero-sum game to guarantee the best outcome under the worst conditions, due to the game's strictly competitive (zero-sum) nature. 

The single-shot game simulation results are obtained from a Monte-Carlo simulation of LTE network and \textit{smart jammer} dynamics as dicussed in Section III. The following parameters are used for our simulations:

\begin{itemize}
\item 	Carrier-to-jammer power ratio: $C/J = 0$ dB,
\item 	Probability of jamming: $p_j = 1.0$,
\item 	Weight of no. of Connected UEs for Type I: $\alpha^{\mathcal{N}_c} = 4$,
\item 	Weight of no. of Connected UEs for Type II: $\alpha^{\mathcal{N}_s} = 5$,
\item 	Weight of average throughput for Type I: $\alpha^{\mathcal{R}_c} = 5$,
\item 	Weight of average throughput for Type II: $\alpha^{\mathcal{R}_{\text{eNB}}} = 4$.
\end{itemize}

The following simulation results are obtained for the single-shot game when the jammer type is \textbf{Cheater}. The first element of the utility matrices represent the \textbf{baseline} scenario when no jammer is active in the network and network is playing its default normal action. 

\[
\mathcal{U}_j^c =
	\begin{bmatrix}
	 -1.0000  & 	-1.0239	&   -2.2464 	&   -1.3840 	&   -1.0000\\
    -0.9642  & -1.0029		&   -2.2130	 	&   -1.3398 	&   -0.9642\\
    -0.8016  & -0.8239 		&  \textbf{-2.0553} 		&   -1.1366 	&   -0.8016\\
    -0.9714  & -1.0078		&   -2.2212  	&  -1.3525 		&   -0.9714\\
    -0.8181  & -0.8399		&   -2.0716  	&  -1.1610 		&   -0.8181
	\end{bmatrix}
\]

Similarly, the simulation results for the single-shot game when the jammer type is \textbf{Saboteur} are presented below.

\[
\mathcal{U}_j^s =
	\begin{bmatrix}
	-1.0000 	&   -0.9933 	&  -0.5635 	&  -0.9128 	&  -1.0000\\
   -0.9879 	&   -0.9805  	&  -0.5446 &  -0.9022	&  -0.9898\\
   -0.9905 	&   -0.9805 	&  -0.4578 &  -0.8849	&  -0.9867\\
   -0.9900 	&   -0.9827 	&  -0.5498 &  -0.9050 &  -0.9919\\
   -0.9895 	&   -0.9800 	&  -0.4666 &  -0.8880	&  -0.9875
	\end{bmatrix}
\]

For the complete information case when the network is aware of the jammer type \textbf{Cheater}, the game has a single \textbf{pure strategy Nash Equilibrium}, $(a^{j \ast},a^{0 \ast})=$ \emph{('Jam CS-RS + PUCCH', 'Throttling')}, with the game value $v = -2.0553$, satisfying the following equation.

\begin{align*} 
  \mathcal{U}_j^c(a^{j*},a^{0*})&=\min_{a^0\in\mathcal{A}_0}\mathcal{U}_j^c(a^{j*},a^0)\\
  &=\max_{a^j\in \mathcal{A}_j}\mathcal{U}_j^c(a^j,a^{0*})
\end{align*}

For the complete information case when the network is aware of the jammer type \textbf{Saboteur}, the game does not have any pure strategy Nash Equilibrium. If the players are allowed to use mixed strategies, i.e., a probability distribution over a player's action set, then there exists a \textbf{mixed strategy Nash Equilibrium} $(x^*,y^*)$, where $x^* = [0\ 0.51\ 0\ 0\ 0.49]^T \in \Delta(\mathcal{A}_j)$, and $y^* = [0.59\ 0\ 0\ 0\ 0.41] \in \Delta(\mathcal{A}_0)$ with the game value $v = -0.9887$,  satisfying the following equation. This mixed strategy probability distribution loosely translates to playing \textit{('Jam CS-RS', 'Jam CS-RS + PCFICH + PUCCH + PRACH')} and \textit{('Normal', 'Timing Change')} equally likely by the jammer and the eNode~B respectively.

\begin{align*}
E_{x^*,y^*}(\mathcal{U}_j^s(a^j,a^0)) 	&=	\min_{y\in\Delta(\mathcal{A}_0)} E_{x^*,y}(\mathcal{U}_j^s(a^j,a^0)) \\
 &= \max_{x\in \Delta(\mathcal{A}_j)}E_{x,y^*}(\mathcal{U}_j^s(a^j,a^0))
\end{align*}

where $E_{x,y}(\mathcal{U}_j^s(a^j,a^0))=x^T \mathcal{U}_j^s y$ is the expected value of the single-stage utility given mixed strategies $x$ and $y$. Given the utility matrix, linear programming is used to compute the Nash Equilibirum \cite{Game_Theory} with $x^*$ and $y^*$, and the game value $v$. 

However, in the asymmetric information  case, eNode B only knows the probability distribution $p_0$ over jammer's types which is public information, while the jammer knows exactly its own type. Knowing its own type, the jammer can use a different strategy for different states $\theta$. Therefore, in the asymmetric game, jammer's mixed strategy $x$ is a mapping from $\Theta$ to $\Delta(\mathcal{A}_j)$. The single-shot asymmetric game still has a mixed strategy Nash Equilibrium $(x^*,y^*)$, where $x^*\in \Delta(\mathcal{A}_j)^{|\Theta|}$ and $y^*\in \Delta(\mathcal{A}_0)$ satisfy

\begin{align*}
E_{p_0,x^*,y^*}(\mathcal{U}_j^\theta(a^j,a^0)) 	&=	\min_{y\in\Delta(\mathcal{A}_0)} E_{p_0,x^*,y}(\mathcal{U}_j^\theta(a^j,a^0)) \\
&=	\max_{x\in \Delta(\mathcal{A}_j)^{|\Theta|}}E_{p_0,x,y^*}(\mathcal{U}_j^\theta(a^j,a^0))
\end{align*}
where $E_{p_0,x,y} \left[ \mathcal{U}_j^\theta(a^j,a^0) \right] = \sum_{\theta \in \Theta} p_0^\theta {x^\theta}^T \mathcal{U}_j^\theta y$ is the expected value of the single-stage utility given the initial probability $p_0$ and mixed strategies $x$ and $y$. It is to be noted here that the utility functions are common knowledge. Although eNode~B is unaware of the jammer type, it knows that the utility function is either $\mathcal{U}_j^c$ or $\mathcal{U}_j^s$ given that the jammer is present in the network. For a given prior probability $p_0$, the eNode~B can  evaluate its expected utility $E_{p_0,x,y} \left[ \mathcal{U}_j^\theta (a^j, a^0) \right]$ whose minmax value (i.e., the game value) is a function of prior probability $p_0$. Since, $p_0$ is fixed, the game value shall also remain fixed. The Nash Equilibrium for the asymmetric information game can be computed by solving an LP by setting the time horizon to a single stage \cite{Sorin_1980}. 

\section{Infinite-Horizon Asymmetric Repeated Game Strategy Algorithms}
The repetition of a zero-sum game in its basic form does not warrant further study as the players can play their optimal security strategies i.i.d. at each stage to guarantee an optimal game value \cite{Zero_Sum}. However, in repeated asymmetric games, playing the optimal strategy in a single-stage asymmetric game i.i.d. at each stage does not guarantee the player the optimal game value \cite{Zero_Sum}. Therefore, the repeated game needs to be studied further.

It is assumed that both players' actions are publicly known at the end of each stage. The jammer's action history at stage $t\geq 1$ is $H^j_t=\{a^j_1,a^j_2,\ldots,a^j_{t-1}\}$, and $\mathcal{H}^j_t$ denotes the set of all possible action histories of the jammer at stage $t$. Similarly, eNode B's action history at stage $t\geq 1$ is $H^0_t=\{a^0_1,a^0_2,\ldots,a^0_{t-1}\}$, and the set of all possible action histories of eNode B at stage $t$ is denoted by $\mathcal{H}^0_t$.

Since the optimal strategies of both players do not depend on the action history of the uninformed player eNode B \cite{Zero_Sum}, the behavior strategy $\sigma_t$ of the jammer at stage $t$ is defined as a mapping from $\Theta\times\mathcal{H}^j_t$ to $\Delta(\mathcal{A}_j)$, and the behavior strategy $\tau_t$ of eNode B at stage $t$ is a mapping from $\mathcal{H}^j_t$ to $\Delta(\mathcal{A}_0)$. The behavior strategies of jammer and eNode B are denoted by  $\sigma=(\sigma_t)_{t=1}^{\infty}$ and $\tau=(\tau_t)_{t=1}^\infty$, and the set of all possible behavior strategies are denoted by $\Sigma$ and $\mathcal{T}$, respectively.

This article considers a $\lambda$-discounted utility function as the overall utility function in the infinite horizon game. The $\lambda$-discounted utility function is commonly used in both classic and modern game-theoretic literature (cf. \cite{Handbook_v1, Incomplete_Information, Zero_Sum, Repeated_Games, Lichun_2015}). The main reason for selecting discounted utility formulation is its guarantee of convergence in infinite-horizon games. If average utility formulation is selected then the overall payoff is taken as a limit, which may or may not exist. Also, uniformity conditions are required for equilibrium in average utility formulations \cite{Handbook_v1}. It is also shown in \cite{Zero_Sum} that as $\lambda \rightarrow 0$, the game value of a discounted infinite-horizon game converges to that of an average reward game. It is to be noted here that $\lambda$ is merely a mathematical constant dictating the discount factor and does not represent any physical quantity like received signal strength or SNR etc. The $\lambda$-discounted utility function is defined as follows:

\begin{equation} \label{eq: discounted_payoff_function}
u_{\lambda}(p_0, \sigma, \tau) 	=		E_{p_0, \sigma, \tau} \left[ \sum_{t=1}^\infty \lambda (1 - \lambda)^{t-1} \mathcal{U}(\theta,  a_t^j, a_t^0) \right]
\end{equation}

Discounted utility represents the idea that players often focus more on current reward and  apply a discount factor of $(1 - \lambda)$ to future reward. Based on the discounted utility function, jammer has a security level $\underline{V}(p_0)=\max_{\sigma\in \Sigma}\min_{\tau\in \mathcal{T}} u_\lambda(p_0,\sigma,\tau)$ which is the maximum utility it can get in the game if eNode B always plays the best response strategy. The strategy $\sigma^*$ that guarantees this value no matter what strategy eNode B plays is called the jammer's security strategy. Similarly, eNode B's security level is defined as $\bar{V}(p_0)=\min_{\tau\in \mathcal{T}} \max_{\sigma\in \Sigma}u_\lambda(p_0,\sigma,\tau)$, and the strategy $\tau^*$ that guarantees the security level no matter what strategy jammer plays is called eNode B's security strategy. If the security levels of both players are the same, which is true in our case, it can be said that the game has a value $V_\lambda(p_0)$, and there exists a Nash Equilibrium.  

This article is concerned  with the security strategies of both players. However, in our case, the system has multiple states and the game is played with the lack of information on one side. Li et al. showed that the security strategies for both the players in finite-horizon asymmetric information repeated zero-sum games depend only on the informed player's history actions \cite{Lichun_2016}. For the infinite-horizon games, this would imply utilizing large amount of memories to record the history actions. It is, therefore, necessary for the players to find fixed-size sufficient statistics for decision making in $\lambda$-discounted infinite-horizon games. However, it is still nontrivial to compute optimal security strategies even with fixed-size sufficient statistics. Therefore, Li \& Shamma provided approximated security strategies with guaranteed performance to solve infinite-horizon games \cite{Lichun_2017}.

\subsection{The Smart Jammer's Approximated Security Strategy Algorithm}
Since jammer's behavior strategy depends on the type of the jammer, the type of the jammer may be revealed through the action history of the jammer. The revelation is characterized by the conditional probability $p_t$ over $\Theta$ conditioned on jammer's action history $H_t^j$, which is updated as follows

\begin{equation} \label{eq: belief_state}
p^{\theta}_{t+1} (H_{t+1}^{j})		= 	\pi(p_t,x_t,a_t^j) 	= 	\frac{p_t^{\theta} (H_t^{j}) x_t^{\theta} (a_t^j)} {\bar{x}_{p_t,x_t} (a_t^j)}		 	
\end{equation}
with $p_1 = p_0$ and $\bar{x}_{p_t,x_t} (a_t^j) = \sum_{\theta \in \Theta} p_t^{\theta} (H_t^{j}) x_t^{\theta} (a_t^j)$ represents weighted average of $x_t$. The conditional probability $p_t$ is also called the belief state which is the sufficient statistics for the informed player, the jammer, to make a decision at stage $t$. It was shown in \cite{Zero_Sum} that the informed player has a stationary security strategy that only depends on $p_t$, and the game value $V_\lambda(p_0)$ satisfies the following recursive equation

\begin{equation} \label{eq: game_value}
\begin{split}
V_\lambda (p_0) 	=		\max_{x \in \Delta (\mathcal{A}_j)^{|\Theta|}} \min_{y \in \Delta (\mathcal{A}_0)}
& \Big[ \lambda \sum_{\theta \in \Theta} p_0^{\theta} {x^{\theta}}^T \mathcal{U}^{\theta} y \\
& + (1 - \lambda) \mathbf{T}_{p_0,x} (V_\lambda) \Big]
\end{split}
\end{equation}
where $x^\theta$ represents jammer's behavioral strategy given state $\theta$, $y$ represents eNode~B's behavioral strategy, and $\mathbf{T}_{p,x} (V_\lambda) = \sum_{a^j \in \mathcal{A}_j} \bar{x}_{p,x} (a^j) V_\lambda (\pi (p, x, a^j))$. Although the  game value $V_{\lambda} (p_t)$ from a stage $t, \forall t \neq 1$ to the end of the game changes as $p_t$ evolves with time, the game value $V_{\lambda} (p_0)$ from the first stage $t = 1$ to the end of the game remains the same as $p_0$ is fixed.

It is non-convex to compute $V_\lambda(p)$ and the corresponding security strategies \cite{Sandholm} and, therefore, an approximated strategy is proposed. The basic idea is to use the game value $V_{\lambda,T}(p)$ of a $T$ stage $\lambda$ discounted asymmetric repeated game to approximate the game value $V_{\lambda,T}(p)$, and compute the approximated security strategy based on the approximated game value $V_{\lambda,T}(p)$. Define the jammer's stationary behavior strategy as $\bar{\sigma}:\Theta\times \Delta(\Theta)\rightarrow \Delta(\mathcal{A}_j)$. The approximated stationary security strategy is 

\begin{equation} \label{eq: J_approx_strategy}
\begin{split}
\bar{\sigma} (:,p) 	=		\argmax_{x \in \Delta (\mathcal{A}_j)^{|\Theta|}}
& \min_{y \in \Delta (\mathcal{A}_0)} \Big[ \lambda \sum_{\theta \in \Theta} p^{\theta} {x^{\theta}}^T \mathcal{U}^{\theta} y \\
& + (1 - \lambda) \mathbf{T}_{p,x} (V_{\lambda,T})   \Big]
\end{split}
\end{equation}
where $\bar{\sigma}(:,p)$ is an $|\mathcal{A}_j| \times |\Theta|$ matrix whose $\theta$ th column is $\bar{\sigma}_{\lambda,T} (\theta,p)$.

Furthermore, Li \& Shamma constructed a linear program to compute the approximated game value $V_{\lambda,T+1} (p)$ and corresponding approximated security strategy $\bar{\sigma}(\theta,p)$. It was shown that $V_{\lambda,T+1} (p)$ satisfies the following linear program in the $\lambda$-discounted  zero-sum asymmetric game $\Gamma_\lambda (p)$:

\begin{equation} \label{eq: J_LP_1}
V_{\lambda,T+1} (p) =	 \max_{q,l} \left[ \sum_{t=1}^{T+1} \sum_{H_t^{j} \in \mathcal{H}_t^{j}} \lambda (1 - \lambda)^{t-1} l_{H_t^{j}}   \right]
\end{equation}

\begin{align}
s.t. \sum_{\theta \in \Theta, a^j \in \mathcal{A}_j} q_{t+1}
& \left( \theta, (H_t^{j}, a^j) \right) \mathcal{U}_{a^j,:}^\theta  \geq   l_{H_t^{j}} \mathbf{1}^T, \\
& \forall t = 1,2,.....,T+1, H_t^{j} \in \mathcal{H}_t^{j}\\
 q_1(H_1^j;\theta)&=1,\forall \theta\in\Theta, \\
 \sum_{a_t^j\in \mathcal{A}_j}q_{t+1}((H_t^j,a_t^j);\theta)&=q_{t}(H_t^j;\theta),\forall \theta\in\Theta, H_{t}^j\in \mathcal{H}_{t}^j,\nonumber\\
  & \forall t=1,\ldots,T,  \\
  q_t(H_t^j;\theta)&\geq 0, \forall \theta\in\Theta, H_{t}^j\in \mathcal{H}_{t}^j,\nonumber\\
& \forall t=2,\ldots,T+1,\label{eq: J_LP_2}
\end{align}
where $q \in Q$ is a set of all properly dimensioned real vectors, $L$ is a properly dimensioned real space, and $(H_t^{j}, a_t^j)$ corresponds to a concatenation. The approximated security strategy is

\begin{equation} \label{eq: J_approx_strategy_solution}
\bar{\sigma}^{a^j} (\theta, p) = q_2^*(a^j;\theta), 	\forall a^j \in \mathcal{A}_j
\end{equation}
where $q_2^*$ is the optimal solution  of the linear program in (\ref{eq: J_LP_1}) - (\ref{eq: J_LP_2}). The curious reader is encouraged to see \cite{Lichun_2017} for further details.

\subsubsection{The Algorithm}
The LP-based algorithm for the informed player to compute its approximated security strategy and update belief state in $\lambda$-discounted asymmetric repeated game is presented as follows \cite{Lichun_2017}:

\textsf{
\begin{enumerate}
\item 	Initialization:
	\begin{enumerate}
	\item 	Read  payoff matrices $\mathcal{U}$, prior probability $p_0$, and system state $\theta$.
	\item 	Set  receding horizon length $T$.
	\item 	Let $t = 1$, and $p_1 = p_0$.
	\end{enumerate}
\item 	Compute the informed player's approximated security strategy $\bar{\sigma}_{\lambda,T}$ based on (\ref{eq: J_approx_strategy_solution}) with $p = p_t$.
\item 	Choose an action $a^j \in \mathcal{A}_j$ according to the probability $\bar{\sigma}_{\lambda,T} (\theta,p_t)$, and announce it publicly.
\item 	Update the belief state $p_{t+1}$ according  to (\ref{eq: belief_state}).
\item 	Update $t = t+1$ and go to step 2.
\end{enumerate}
}

\subsection{The eNode~B's Approximated Security Strategy Algorithm}
The uninformed player does not have access to the informed player's strategy or belief state $p_t$, therefore, $p_t$ cannot serve as its sufficient statistics. The sufficient statistics of the uninformed player, eNode B, was shown to be the anti-discounted regret which will be explained further. The regret $\mu_t^\theta(H_t^j)$ in state $\theta$ is defined as the difference between the expected realized utility so far and the security level of eNode B's security strategy, given state $\theta$, i.e.,

\begin{align*}
  \mu_t^\theta(H_t^j)=E_{\tau}\left(\sum_{s=1}^{t-1}\lambda(1-\lambda)^{s-1}\mathcal{U}^j(a^j_s,a^0_s)|\theta,H_t^j\right) -\mu^{\theta*}
\end{align*}
where $$\mu^{\theta*}=\max_{\sigma(\theta)\in \Sigma(\theta)}E_{\sigma(\theta),\tau^*}\left(\sum_{s=1}^\infty \lambda(1-\lambda)^{s-1}\mathcal{U}^j(a^j_s,a^0_s)|\theta\right),$$
 $\tau^*$ is eNode B's security strategy, $\sigma(\theta)$ indicates jammer's behavior strategy given $\theta\in \Theta$, and $\Sigma(\theta)$ is the corresponding set including all $\sigma(\theta)$. The anti-discounted regret is defined as

\begin{align*}
  w_t^\theta(H_t^j)=\frac{\mu_t^\theta(H_t^j)}{(1-\lambda)^{t-1}},\forall \theta\in \Theta,
\end{align*}

and is updated according to

\begin{align} \label{eq: anti_discounted_computation}
  w_{t+1}^\theta(H_t^j,a^j_t)=\frac{w_t^\theta(H_t^j)+\lambda\mathcal{U}^\theta(a^j_t,:)\tau(H_t^j)}{1-\lambda}, \forall \theta\in \Theta, 
\end{align}
where $\mathcal{U}^\theta(a^j_t,:)$ is the $a^j_t$th row of matrix $\mathcal{U}^\theta$.

Computation of the security level $\mu^*$ of eNode B's security strategy is non-convex \cite{Lichun_2017}. Therefore, an approximated security level $\mu^{\theta\star}$ is used, which is the security level of eNode~B's security strategy given state $\theta$ in $T$-stage $\lambda$-discounted asymmetric repeated game. The approximated security level $\mu^{\theta\star}$ is computed according to the following linear program:

\begin{equation} \label{eq: primal_game_LP_1}
\min_{y \in Y, l \in \mathbb{R}^{|\Theta|}} \sum_{\theta \in \Theta} p^\theta l^\theta
\end{equation}

\begin{align}
s.t. \; &\sum_{t=1}^T \lambda (1 - \lambda)^{t-1} \mathcal{U}^\theta(a_t^j,:) y_{H_{T}^{j}} 	\leq 	l^\theta , \forall \theta \in \Theta,\nonumber\\
&\forall H_{T}^j\in\mathcal{H}^j_{T},a_t^j\in \mathcal{A}_j,\label{eq: primal_game_LP_2}
\end{align}

\begin{equation} \label{eq: primal_game_LP_3}
\mathbf{1}^T y_{H_t^{j}} = 1, \forall H_t^{j} \in \mathcal{H}_t^{j}, \forall t = 1,....,T,
\end{equation}

\begin{equation} \label{eq: primal_game_LP_4}
y_{H_t^{j}} 	\geq  	\mathbf{0}, \forall H_t^{j}\in \mathcal{H}_t^{j}, \forall t = 1,....,T
\end{equation}
where $Y$ is properly-dimensioned real space. The approximated security level is $\mu^{\theta\star} = l^*$, where $l^*$ is the optimal solution to the LP problem (\ref{eq: primal_game_LP_1}-\ref{eq: primal_game_LP_4}).

The eNode B has a stationary security strategy that only depends on the anti-discounted regret $w_t$ \cite{Lichun_2017}. Define eNode B's stationary behavior strategy as $\bar{\tau}:\mathbb{R}^{|\theta|}\rightarrow \Delta(\mathcal{A}_0)$. Computation of the stationary security strategy of eNode B is non-convex \cite{Lichun_2017}. Therefore, an approximated stationary security strategy $\bar{\tau}(w)$ of eNode B is proposed in \cite{Lichun_2017}, which can be computed by solving the following LP problem.

\begin{equation} \label{eq: dual_game_LP_1}
\min_{y \in Y, l \in \mathbb{R}^{|\Theta|}, L \in \mathbb{R}} L
\end{equation}

\begin{equation} \label{eq: dual_game_LP_2}
s.t. \; w + l 	\leq  	L \mathbf{1},
\end{equation}

\begin{align}
s.t. \; &\sum_{t=1}^{T+1} \lambda (1 - \lambda)^{t-1} \mathcal{U}^\theta(a^j,:) y_{H_{T+1}^{j}} 	\leq 	l^\theta , \forall \theta \in \Theta,\nonumber\\
&\forall H_{T+1}^j\in\mathcal{H}^j_{T+1},a^j\in \mathcal{A}_j,\label{eq: dual_game_LP_3}
\end{align}

\begin{equation} \label{eq: dual_game_LP_4}
\mathbf{1}^T y_{H_t^{_j}} = 1, \forall H_t^{j} \in \mathcal{H}_t^{j}, \forall t = 1,....,T+1,
\end{equation}

\begin{equation} \label{eq: dual_game_LP_5}
y_{H_t^{j}} 	\geq  	\mathbf{0}, \forall H_t^{j} \in \mathcal{H}_t^{j}, \forall t = 1,....,T+1
\end{equation}
where $Y$ is properly-dimensioned real space. The uninformed player's approximated security strategy $\bar{\tau}(w)$ is $y_{H_1^{j}}^*$. The curious reader is encouraged to see \cite{Lichun_2017} for further details.

\subsubsection{The Algorithm}
The LP-based algorithm for the uninformed player to compute its approximated security strategy in $\lambda$-discounted asymmetric repeated game $\Gamma_\lambda (p_0)$ is presented as follows \cite{Lichun_2017}:

\textsf{
\begin{enumerate}
\item 	Initialization:
	\begin{enumerate}
	\item 	Read payoff matrices $\mathcal{U}$, and prior probability $p_0$.
	\item 	Set receding horizon length $T$.
	\item 	Solve the LP problem (\ref{eq: primal_game_LP_1}) - (\ref{eq: primal_game_LP_4}) with $p = p_0$ and let $\mu^\ast = l^\ast$.
	\item 	Let $t = 1$ and $w_1 =- \mu^\ast$.
	\end{enumerate}
\item 	Solve the LP problem (\ref{eq: dual_game_LP_1}) - (\ref{eq: dual_game_LP_5}) with $w = w_t$, and the uninformed player's approximated security strategy $\bar{\tau}(w_t)$ is $y_{H_1^{j}}^\ast$.
\item 	Choose an action $a^0 \in \mathcal{A}_0$ according to the probability $\bar{\tau}(w_t)$, and announce it publicly.
\item 	Read the informed player's action, and update the anti-discounted regret $w_{t+1}$ according to (\ref{eq: anti_discounted_computation}).
\item 	Update $t = t+1$ and  go to Step 2.
\end{enumerate}
}

It is to be noted here that jamming sense is still required by the network,  i.e., the network has to first decide whether it is under jamming attack or not in order to invoke the above-mentioned algorithm. 

\subsection{The eNode~B's Expected Security Strategy Algorithm}
The ``expected strategy'' algorithm for the eNode~B is defined as a simplex $\Delta$ over its complete-information single-shot game security strategies $\sigma_1$ with the same probability as prior $p_0$. In other words, the eNode~B would play the complete-information single-shot security strategies $\sigma_1|\theta = 1$ and $\sigma_1|\theta = 2$ with the probabilities $p_0^1$ and $p_0^2$ respectively. Since, the prior is common knowledge, it alleviates eNode~B from ``learning'' and full monitoring in a repeated game. Thus, the eNode~B essentially plays a single-shot strategy in a repeated game but without the requirements of full monitoring, which may not be such a bad idea if the jammer plays ``non-revealing'' strategies. Furthermore, the network does not need to observe the jammer's action with certainty that leads to more practical implementations. Both discounted and average payoff formulations can be used with this algorithm. It is to be noted here that the expected security strategy algorithm is novel and not based on a prior work.

In the next section, the approximated security strategy and expected security strategy algorithms are used to design strategies for both the \textit{smart jammer} and the LTE network. The $\lambda$-discounted cost formulation is used for both the algorithms in the infinite-horizon game.

\section{Performance Analysis of Repeated Game Strategy Algorithms}
The zero-sum game-theoretic algorithms presented earlier are used to devise ``approximated'' strategy formulations for both the players in  a $\lambda$-discounted utility sense. However, these algorithms require full monitoring, i.e. the network has to observe jammer's action at every stage with certainty. Therefore, the ``expected'' formulation is devised in which the network being the uninformed player simply plays its single-shot \textit{best response} in an expected sense, i.e., it would play  single-shot Best Response (BR) with the same  probability distribution as the prior probability (which is common knowledge) of the jammer occurrence. This enables  the network to alleviate full monitoring requirement, i.e., the network does not have to observe the jammer's action with certainty and leads to more practical implementations.

The performance of both ``approximated'' and ``expected'' algorithms for discounted utility formulations is characterized in the following section. However, not all of the simulation results can be shared here due to space constraints. The following parameters were used for both the players in repeated game simulations (in addition to the single-shot case): discount factor $\lambda = 0.90$ and receding horizon length $T = 4$. It is to be noted here that the receding horizon length of $T = 4$ is chosen for simulation efficiency purposes and almost the same results are obtained at higher values of T.

\subsection{eNode~B vs. Cheater}
\subsubsection{Jammer Strategy}
\begin{figure}
\centering
\includegraphics[width= 0.5 \textwidth]{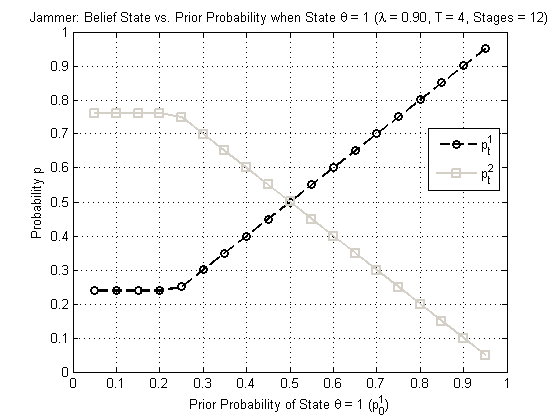}
\caption{Cheater's Steady State Belief State vs. Prior $p_0^1$}
\label{fig: Cheater_belief_state}
\end{figure}

\begin{figure}
\centering
\includegraphics[width= 0.5  \textwidth]{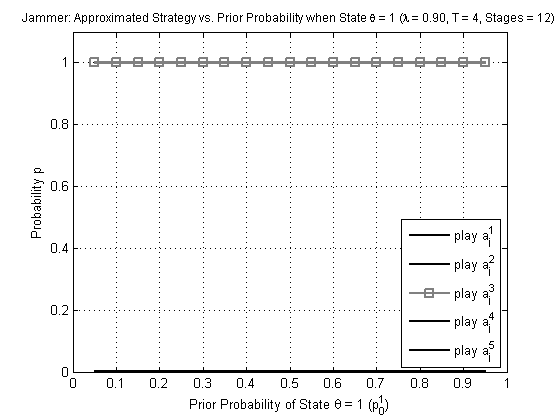}
\caption{Cheater's Steady State Approximated Security Strategy vs. Prior $p_0^1$}
\label{fig: Cheater_approximated_strategy}
\end{figure}

\begin{figure}
\centering
\includegraphics[width= 0.5 \textwidth]{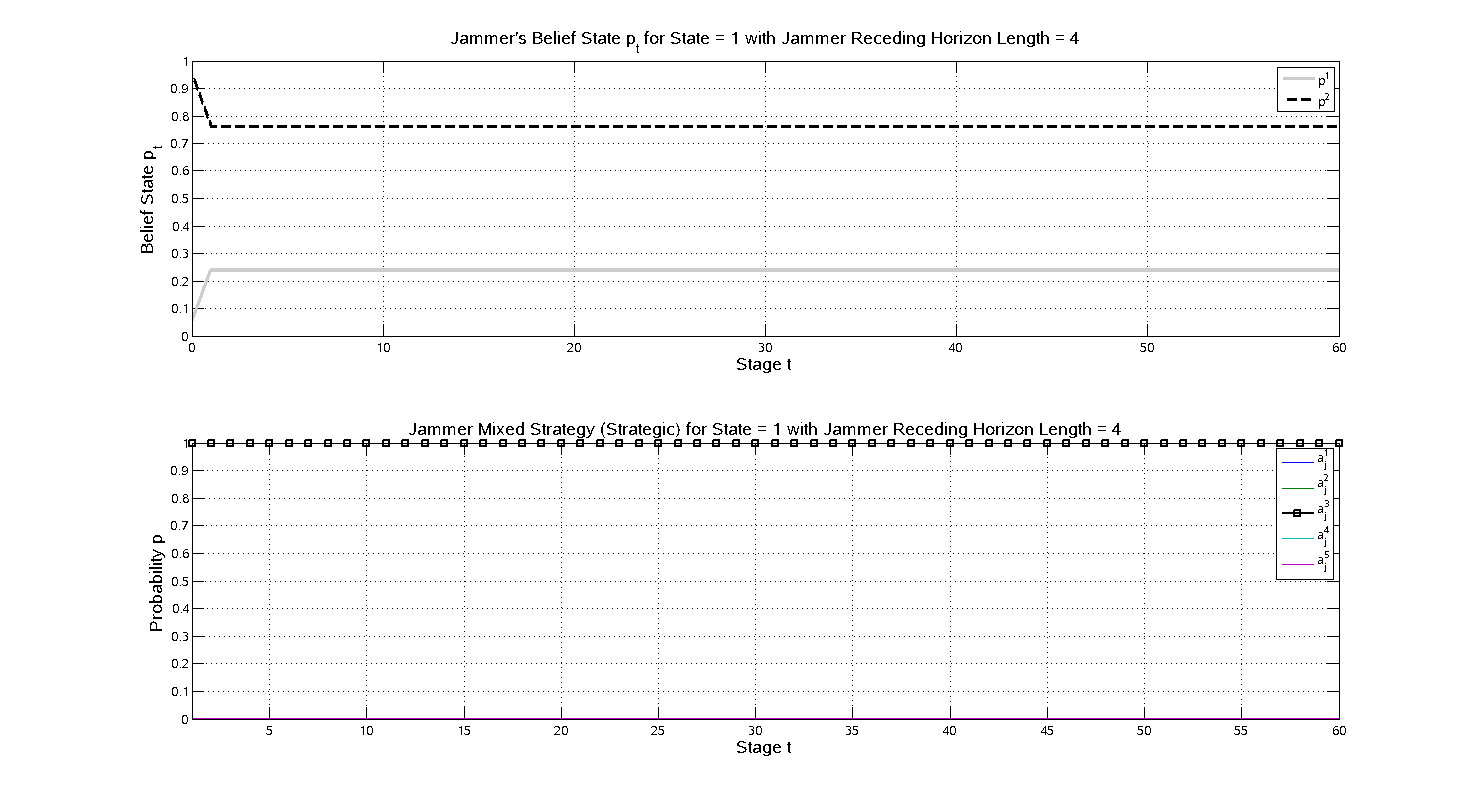}
\caption{Cheater's Belief \& Strategy vs. Time when $p_0^1 = 0.05$}
\label{fig: cheater_strategy_vs_time}
\end{figure}

When the \textit{Cheater} ($\theta = 1$) is in the network, it always uses its ``approximated'' algorithm to devise repeated game strategy against the network. Also, being the informed player, there is no  ambiguity about the system state so \textit{Cheater} can decide to reveal its superior information as much as it suits it. The Cheater's steady state belief state $p_t$ and repeated game strategy vs. prior probability are shown in Figs. \ref{fig: Cheater_belief_state} and \ref{fig: Cheater_approximated_strategy}, respectively, where $p^1$ and $p^2$ represent updated belief (probability) about the states $\theta = 1$ and $\theta = 2$, respectively, and $a_j^k$ represents \textit{kth} pure action of the Cheater. It is interesting to note that the Cheater always plays the same security (pure) strategy (play $a_j^3$ = \textit{'Jam CS-RS + PUCCH'}) that it uses for a single-shot game, independent of the prior probability. It is also interesting to know that Cheater's strategies are \textbf{non-revealing}~\footnote{The informed player is said to play non-revealing at stage $n$ when the posterior probabilities in (\ref{eq: belief_state}) do not change at that stage if its mixed move at  stage $n$ is independent of the state $\theta \in \Theta$ for all values of $\theta$ for which  $p_n^\theta > 0$. In case when full monitoring is assumed, not revealing the information is equivalent to not using that information, \cite{Repeated_Games}.}, even at a relatively low prior probability of its occurrence when $p_0^1 \geq 0.25$. This means that the network does not ``learn'' anything new about the jammer type from jammer's repeated actions despite full monitoring when $p_0^1 \geq 0.25$ and Cheater takes full advantage  of its superior information. At relatively low prior probability of Cheater's occurrence ($p_0^1 < 0.25$), the jammer reveals very little information in the first stage when the belief state gets updated to $p_0 = [0.25 \; 0.75]^T,$ but it remains the same after that. For instance, Fig. \ref{fig: cheater_strategy_vs_time} shows the evoluation of Cheater's belief state and its strategy at every stage when the prior probability is $p_0^1 = 0.05$. This puts the network at a disadvantageous position in the game if the network plays as a Bayesian player, even when it can observe jammer's actions perfectly at every stage.

\subsubsection{eNode~B Strategies}
\begin{figure}
\centering
\includegraphics[width= 0.5 \textwidth]{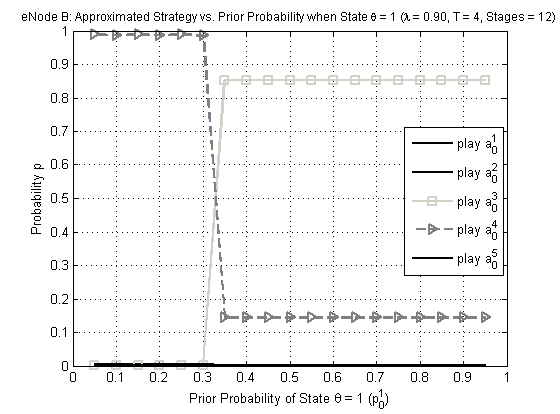}
\caption{eNode~B's Steady State Approximated Security Strategy against Cheater vs. Prior $p_0^1$}
\label{fig: eNB_approximated_strategy_state_1}
\end{figure}

\begin{figure}
\centering
\includegraphics[width= 0.5 \textwidth]{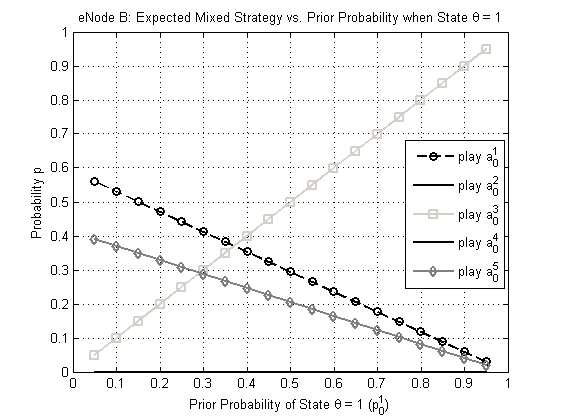}
\caption{eNode~B's Expected Strategy vs. Prior $p_0^1$}
\label{fig: eNB_expected_strategy_state_1}
\end{figure}

The eNode~B's steady state ``approximated'' and ``expected'' security strategies vs. prior probability $p_0^1$ are plotted in Figs. \ref{fig: eNB_approximated_strategy_state_1}, and \ref{fig: eNB_expected_strategy_state_1}, respectively, where $a_0^k$ represents \textit{kth} pure action of the eNode~B. The network's strategies (both ``expected'' and ``approximated'') evolve with varying prior probability levels as it is the uninformed player. The ``approximated'' strategy relies on full monitoring and switches to a different strategy at $p_0^1 \geq 0.35$, when it starts playing $a_0^3$ = \textit{`Throttling'} (its security strategy against Cheater in complete-information single-shot game) in addition to playing $a_0^4$ =  \textit{`Change~$f_c$'}. On the other hand, the ``expected'' algorithm does not rely on full monitoring and, hence, uses an expectation of its single-shot strategies involving playing mixed strategy over \textit{`Normal', `Throttling'} and \textit{`Change Timing'}. The ``expected'' strategy is pre-computed based on the prior probability and does not change as the game  proceeds, whereas  the ``approximated'' algorithm converges in around 12 stages. The ``expected'' strategy algorithm may work well enough for the network as the jammer's strategies are mostly non-revealing and the  ``approximated'' algorithm requires full monitoring.

\subsubsection{eNode~B's $\lambda$-discounted Utilities}
\begin{figure}
\centering
\includegraphics[width= 0.5 \textwidth]{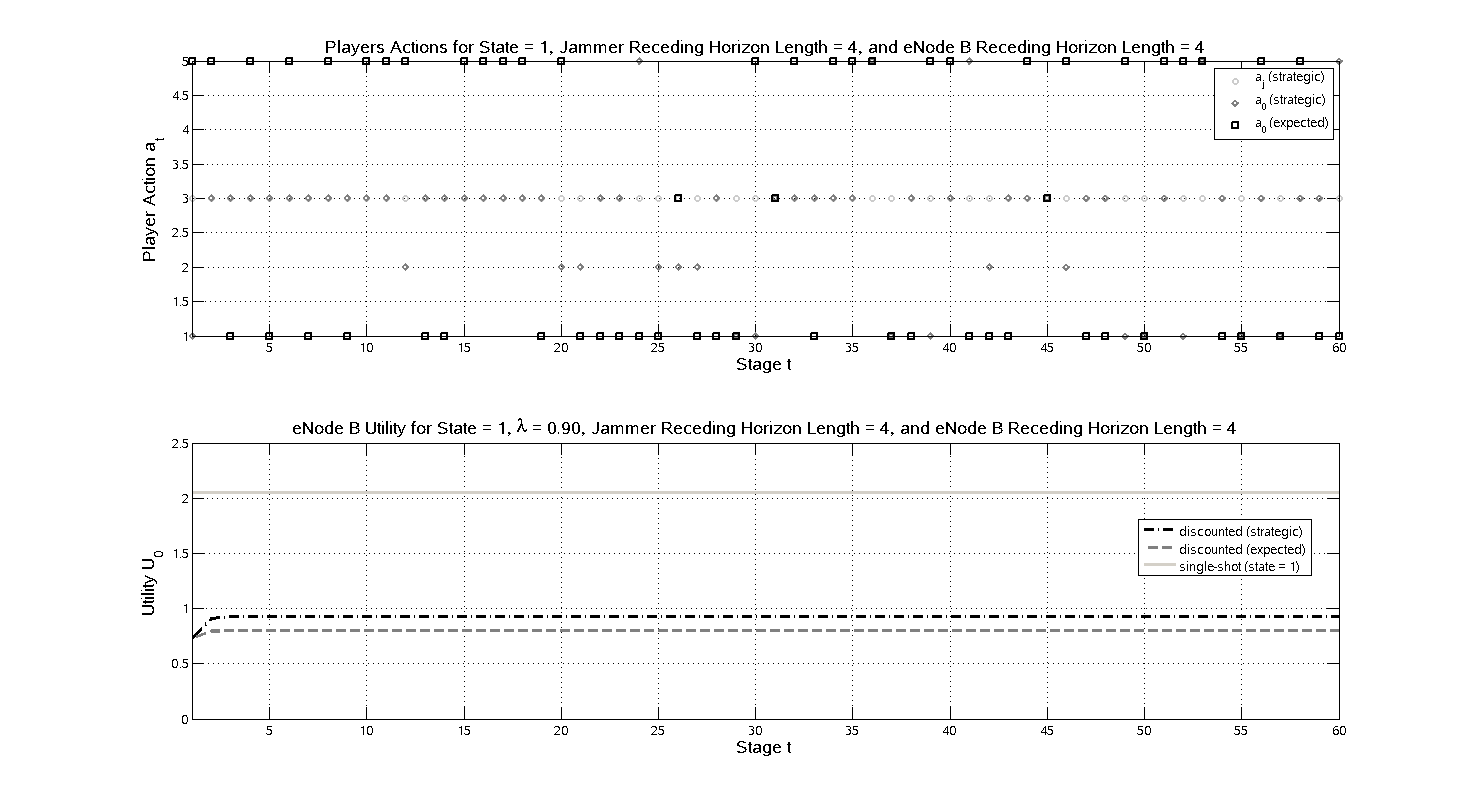}
\caption{Players' Actions \& Utility vs. Time when $p_0^1 = 0.05$}
\label{fig: eNB_utility_state_1_vs_time}
\end{figure}

\begin{figure}
\centering
\includegraphics[width= 0.5 \textwidth]{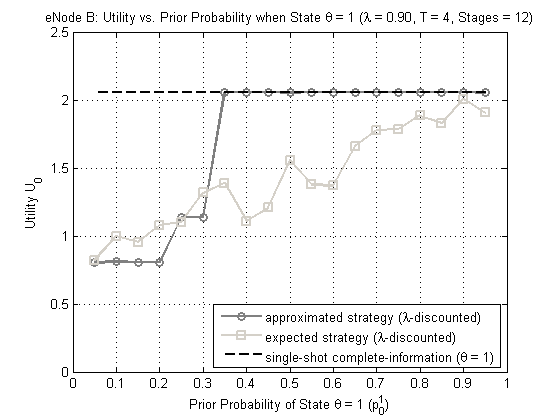}
\caption{eNode~B's Utility against Cheater vs. Prior $p_0^1$}
\label{fig: eNB_utility_state_1}
\end{figure}

A snapshot of both eNode~B and Cheater's actions and eNode~B's utility at every stage is shown in Fig. \ref{fig: eNB_utility_state_1_vs_time} for $\lambda = 0.90$ and $p_0^1 = 0.05$. It is apparent that the eNode~B's (hence, Cheater's) utility stabilizes very quickly at the beginning of the game - a trend that is observed throughout the repeated game. The eNode~B's ``approximated'' and ``expected'' $\lambda$-discounted utility values against Cheater are plotted in Fig. \ref{fig: eNB_utility_state_1} at different prior probability $p_0^1$ levels. The ``approximated security'' algorithm performs almost optimally when $p_0^1 \geq 0.35$, whereas the ``expected'' algorithm performs poorly as compared to the ``approximated'' algorithm with the exception of low prior values. The ``approximated'' algorithm uses full monitoring and repeated game linear programming (LP) formulation to compute its strategy and, hence, performs much better than  its counterpart. On the other hand, the ``expected'' algorithm only relies on the prior probability and does not observe the jammer's actions and, hence, ends up underperforming even when the jammer uses its single-shot security strategy. When prior probability for Cheater's occurrence is low (i.e., $p_0^1 < 0.35$), eNode~B strategies fail to even come close to the complete-information single-shot value. This happens due to the fact that it is rather unlikely for the Cheater to be present in the network at such low prior value and eNode~B strategy algorithms are not robust enough to address this problem.

\subsection{eNode~B vs. Saboteur}
\subsubsection{Jammer Strategy}
\begin{figure}
\centering
\includegraphics[width= 0.5 \textwidth]{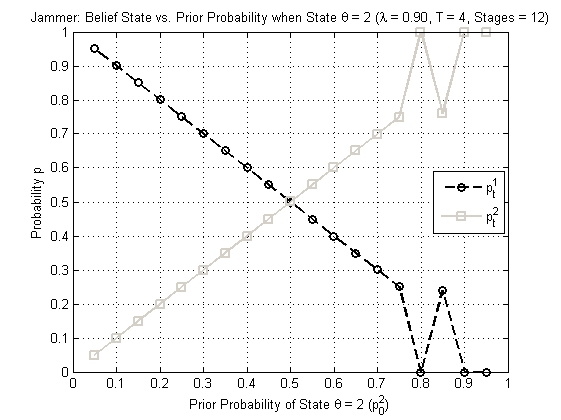}
\caption{Saboteur's Steady State Belief State vs. Prior $p_0^2$}
\label{fig: Saboteur_belief_state}
\end{figure}

\begin{figure}
\centering
\includegraphics[width= 0.5  \textwidth]{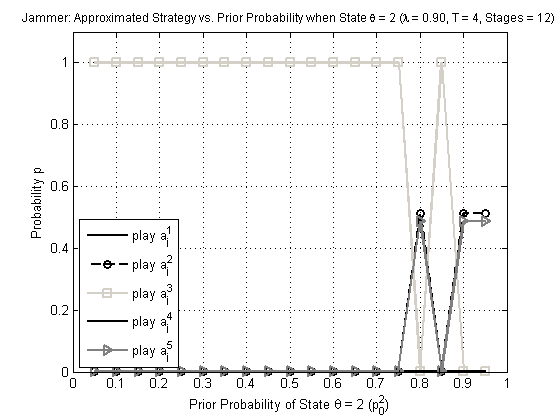}
\caption{Saboteur's Steady State Approximated Security Strategy vs. Prior $p_0^2$}
\label{fig: Saboteur_approximated_strategy}
\end{figure}

\begin{figure}
\centering
\includegraphics[width= 0.5 \textwidth]{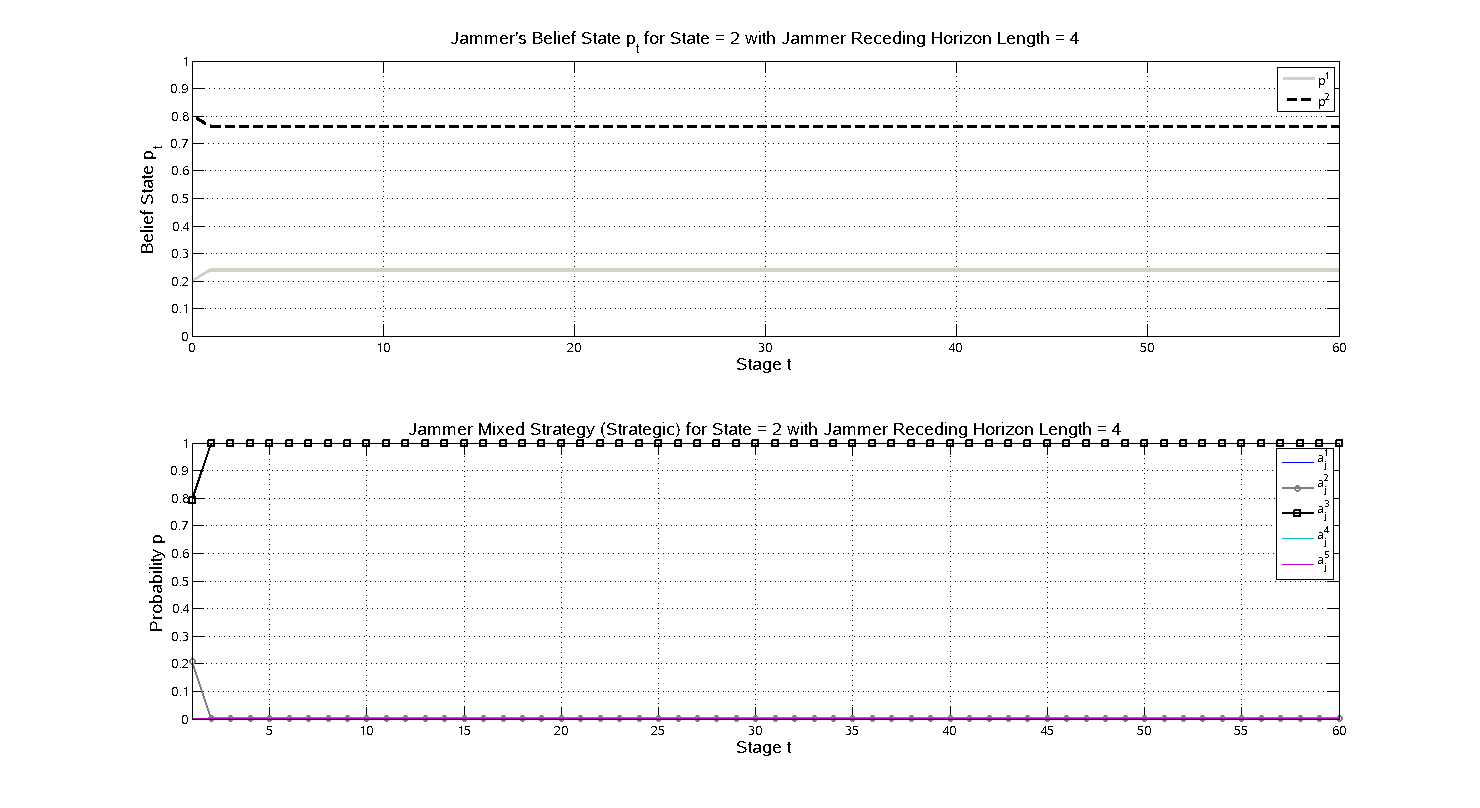}
\caption{Saboteur's Belief \& Strategy vs. Time when $p_0^2 = 0.80$}
\label{fig: Saboteur_strategy_vs_time}
\end{figure}

Similar to the eNode~B vs. Cheater game, Saboteur's steady state belief states $p_t$ and ``approximated security'' strategies vs. prior probability of its occurrence $p_0^2$ are shown in Figs. \ref{fig: Saboteur_belief_state} and \ref{fig: Saboteur_approximated_strategy}, respectively. It is very interesting to note that being the informed player, Saboteur plays \textbf{non-revealing} and \textbf{``misleading''} strategies even at prior probability values as high as $p_0^2 = 0.75$ (this value goes up to $p_0^2 = 0.85$ for $\lambda = 0.70$). It plays its type $\theta  = 1$ (Cheater) dominant security strategy (play $a_j^3$ = \textit{'Jam CS-RS + PUCCH'}) while actually being  a  type $\theta = 2$ (Saboteur) jammer. For example, Fig. \ref{fig: Saboteur_strategy_vs_time} shows the evoluation of Saboteur's belief state and its strategy at every stage when the prior probability is $p_0^2 = 0.80$. At high prior probability values of $0.75 < p_0^2 < 0.90$, the jammer's belief state goes through a transition period because the network forces it to reveal its true identity by playing $a_0^4$ with certainty. Hence, the belief state eventually settles down to the completely-revealing state of $p^2 = 1$. During the transition period when the jammer's belief state converges to $p^2 = 1$, it plays its single-shot security strategy for state $\theta = 2$, i.e., play $a_j^2$ = \textit{'Jam CS-RS'} and $a_j^5$ = \textit{'Jam CS-RS + PUCCH + PCFICH + PRACH'} with almost the same probability. At very high prior probability levels of $p_0^2 \geq 0.90$, the state information ($\theta = 2$) is completely revealed and the jammer plays its single-shot security strategy for state $\theta = 2$ as mentioned above. Hence, the jammer uses its superior information to its complete advantage even when full monitoring is  allowed. This is a good example of the strength of superior information and how it can be exploited in asymmetric games against an adversary.

\subsubsection{eNode~B Strategies}
\begin{figure}
\centering
\includegraphics[width= 0.5 \textwidth]{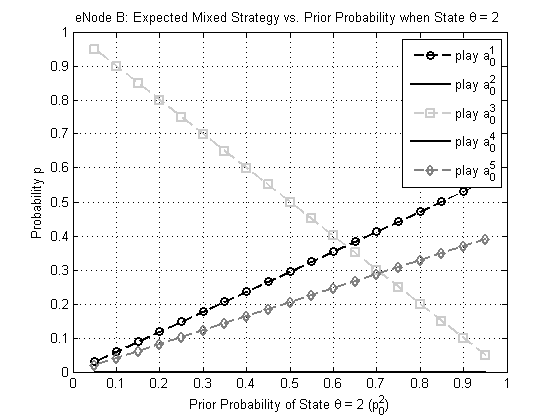}
\caption{eNode~B's Expected Strategy vs. Prior $p_0^2$}
\label{fig: eNB_expected_strategy_state_2}
\end{figure}

Similar to the repeated game against Cheater, the eNode~B adapts its repeated game strategy against Saboteur as the game proceeds. From the simulations, eNode~B's strategy seems to converge in 12 stages. The ``expected'' strategy is shown in Fig. \ref{fig: eNB_expected_strategy_state_2} and is deployed similar to the game against Cheater. Since, the ``expected'' strategy algorithm is oblivious to the actual jammer type and does not use full monitoring, its mixed strategy does not depend on the system state $\theta$ and is played solely based on the prior probability value. 

\begin{figure}
\centering
\includegraphics[width= 0.5 \textwidth]{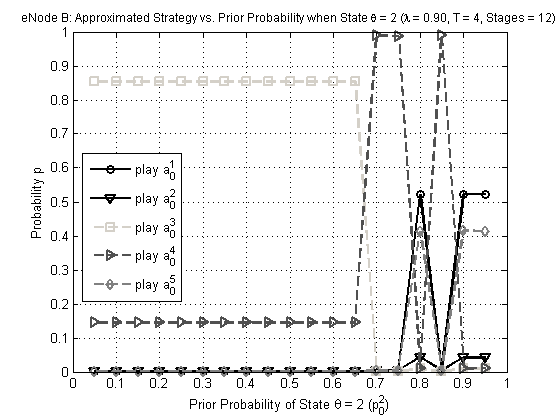}
\caption{eNode~B's Steady State Approximated Security Strategy against Saboteur vs. Prior $p_0^2$}
\label{fig: eNB_approximated_strategy_state_2}
\end{figure}

On the other hand, the ``approximated'' security strategy algorithm relies on the repeated game and full monitoring to adapt its strategy. The network's steady state ``approximated'' strategy vs. prior probability $p_0^2$ is plotted in Fig. \ref{fig: eNB_approximated_strategy_state_2}. As discussed above, the jammer plays completely non-revealing and misleading strategies for $p_0^2 \leq 0.75$ and, hence, eNode~B gets tricked into believing that it is playing against Cheater ($\theta = 1$), when in fact it is playing against the Saboteur ($\theta  = 2$). This leads the network to play the same strategy that it played against Cheater until $p_0^2 \leq 0.70$. At $0.70 \leq p_0^2 \leq 0.75$, the network plays $a_0^4$ = \textit{'Change frequency'} with certainty to force the jammer to reveal its state. When the jammer starts playing  its security strategy for state $\theta = 2$ at high prior values, the network switches to its own security strategy against Saboteur and plays $a_0^1$ = \textit{'Normal'} + $a_0^5$ = \textit{'Change Timing'}. The network also plays $a_0^2$ = \textit{'Pilot Boosting'} with a very low probability. This trend continues whenever the network observes (due to full monitoring) the jammer playing its state $\theta = 2$ security strategies. It is curious to see how the network gets tricked by the jammer even  with full monitoring because it lacks information about the system state.

\subsubsection{eNode~B's $\lambda$-discounted Utilities}
\begin{figure}
\centering
\includegraphics[width= 0.5 \textwidth]{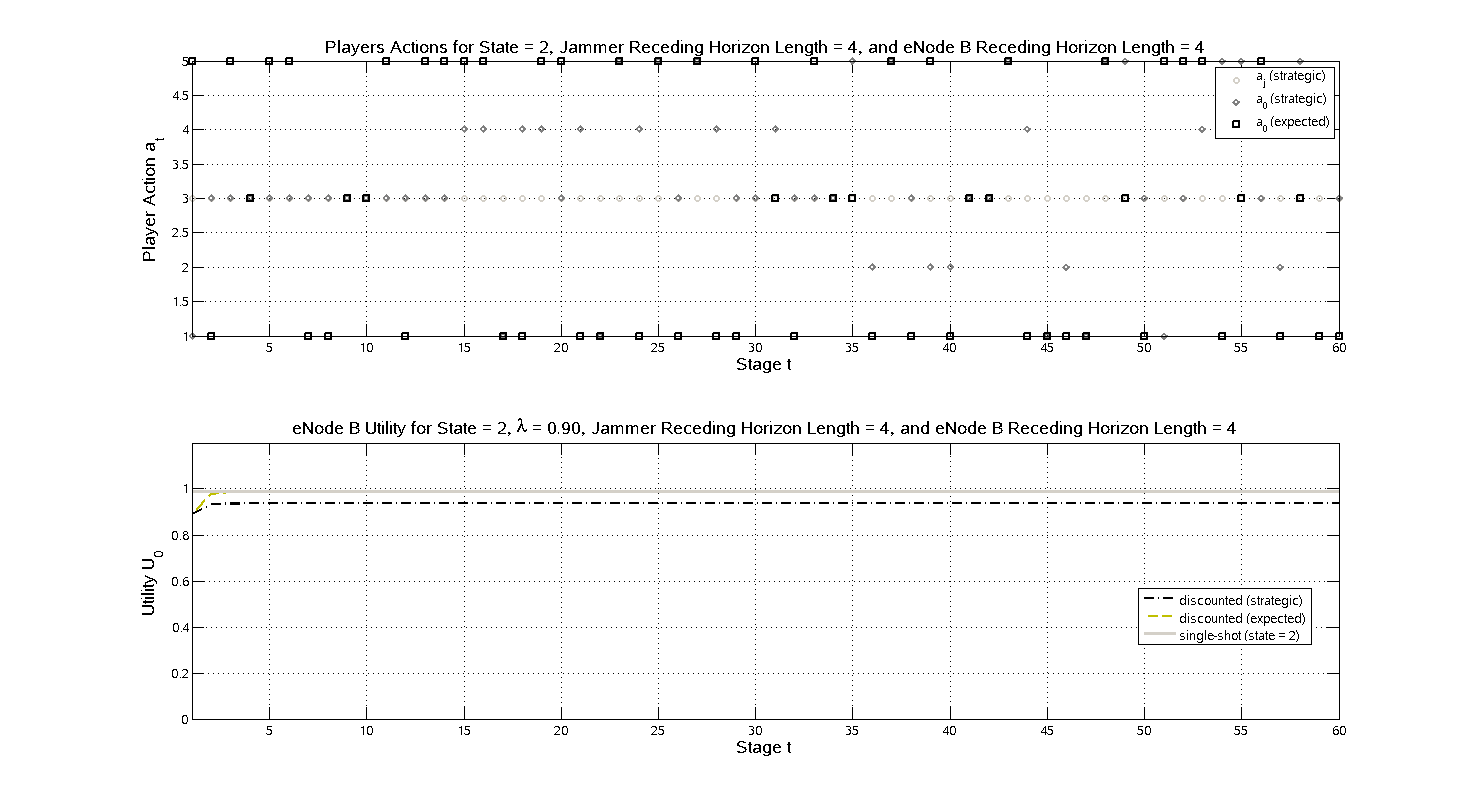}
\caption{Players' Actions \& Utility vs. Time when $p_0^2 = 0.80$}
\label{fig: eNB_utility_state_2_vs_time}
\end{figure}

\begin{figure}
\centering
\includegraphics[width= 0.5 \textwidth]{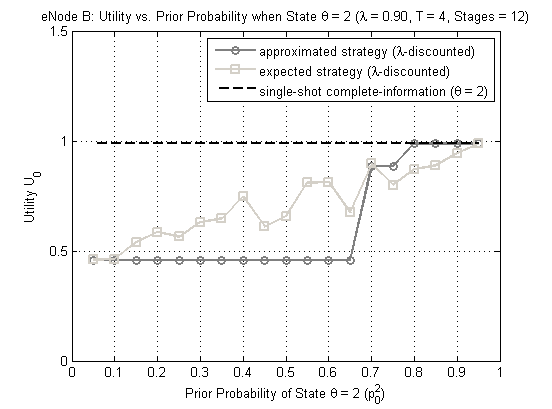}
\caption{eNode~B's Utility against Saboteur vs. Prior $p_0^2$}
\label{fig: eNB_utility_state_2}
\end{figure}

A snapshot of both eNode~B and Saboteur's actions and eNode~B's utility at every stage is shown in Fig. \ref{fig: eNB_utility_state_2_vs_time} for $\lambda = 0.90$ and $p_0^2 = 0.80$. Similar to the game against Cheater, eNode~B's (hence, Saboteur's) utility stabilizes very quickly at the beginning of the game. The network's $\lambda$-discounted utility values for both ``approximated'' and ``expected'' security strategy algorithms are plotted against prior probability $p_0^2$ in Fig. \ref{fig: eNB_utility_state_2}. The jammer strategies are mostly non-revealing and, hence, eNode~B does not seem to ``learn'' much about the jammer type from its repeated interaction. Therefore, the ``approximated security strategy'' formulation seems to perform very poorly until $p_0^2 < 0.70$. At $p_0^2 = 0.70$, the eNode~B switches its strategy to playing $a_0^4$ = \textit{'Change  $f_c$'} and catches up to the optimal value at $p_0^2 \geq 0.80$. Obviously, the jammer also uses full monitoring and is forced to come out and play revealing strategy at $p_0^2 \geq 0.90$. 

On the other hand, the ``expected strategy'' algorithm seems to perform better than the ``approximated security strategy'' as it does not get tricked by the jammer's non-revealing strategies due to its oblivion. It appears that the ``expected strategy'' algorithm outperforms its counterpart when $p_0^1 \leq 0.30$ (or equivalently, $p_0^2 \geq 0.70$) given  that the Cheater ($\theta = 1$) is present in the network and $p_0^2 \leq 0.70$ (or equivalently, $p_0^1 \geq 0.30$) when Saboteur ($\theta = 2$) is present in the network. Thus, it performs better in low prior probability regions, when eNode~B does not expect a certain jammer type in the network. 

Nevertheless, it becomes clear that the network is at a very disadvantageous position in the game against the \textit{smart jammer} due to its lack of information and can be easily misled by the jammer. Furthermore, the ``approximated'' and ``expected'' strategy algorithms work in a complementary sense in favor of the network.

\section{Conclusion}
In this article, the smart jammer and eNode~B dynamics are modeled as a strictly competitive (zero-sum) repeated asymmetric game with incomplete information and lack of information on the network side. The solution of a complete-information single-shot game is based on very familiar security strategies that lead to a Nash equilibrium. However, tractable optimal strategy formulations for infinite-horizon asymmetric repeated games do not exist in game-theoretic literature, especially for the uninformed player. Therefore, efficient LP formulations from a recent work  are used  for ``approximated'' security strategy computation for  both players that requires full monitoring. Therefore, a simplistic yet effective ``expected'' security strategy algorithm is also devised for the network that does not require full monitoring. 

This article also presents and discusses performance characterization of the above-mentioned algorithms. It turns out that the jammer is able to play non-revealing strategies most of the time, which implies that the network is unable to learn any new information about the jammer type in repeated games. Hence, at low prior values, the network performs worse (or equivalently, smart jammer performs better) in repeated games as compared to the complete-information single-shot game. In the  game against  the Cheater, the ``approximated security strategy'' algorithm is able to strategize against the Cheater rather quickly and achieves its optimal utility because the jammer plays its single-shot game security strategy in repeated game. However, this advantage goes away in the game against the Saboteur, when  the jammer plays misleading strategies for a wide range of prior probabilities. Nevertheless, the network's algorithm eventually catches up and forces the jammer to reveal its true type.

The unique ``expected security strategy'' algorithm performs equally well or sometimes better than its counterpart ``approximated security strategy'' algorithm against the type Saboteur. This is due to the fact that it does not get duped by the ``misinformation'' spread by the jammer due to lack of full monitoring, which plays at its advantage. However, the former algorithm performs better than the latter at low prior probability values against the type Cheater because the smart jammer always plays its single-shot game security strategy. Nevertheless, the biggest advantage of ``expected strategy'' algorithm comes from the fact that it does not require full monitoring and, hence, can be easily deployed in practical networks. 


\ifCLASSOPTIONcaptionsoff
  \newpage
\fi

\end{document}